\documentclass{article}

\PassOptionsToPackage{numbers, compress}{natbib}
\usepackage[preprint]{neurips_2026}
\usepackage{graphicx}
\usepackage{amssymb}
\usepackage{multirow}
\usepackage{amsmath}
\usepackage{amsfonts}

\usepackage[utf8]{inputenc} 
\usepackage[T1]{fontenc}    
\usepackage{hyperref}       
\usepackage{url}            
\usepackage{booktabs}       
\usepackage{amsfonts}       
\usepackage{nicefrac}       
\usepackage{microtype}      
\usepackage{xcolor}         

\title{Modality-Aware Contrastive and Uncertainty-Regularized Emotion Recognition}

\author{Yan Zhuang$^{1}$, Minhao Liu$^{1,2}$, Yanru Zhang$^{1,2}$, Jiawen Deng$^{1,*}$, Fuji Ren$^{1,2,*}$\\
$^1$University of Electronic Science and Technology of China \\ 
$^2$Shenzhen Institute for Advanced Study, UESTC \\
{\tt\small \{202211081370\}@std.uestc.edu.cn} \\
{\tt\small \{minhaoliu,yanruzhang,dengjw,renfuji\}@uestc.edu.cn} \\
$^*$Corresponding authors
}

\begin{document}

\maketitle

\begin{abstract}
Multimodal Emotion Recognition (MER) has attracted growing attention with the rapid advancement of human–computer interaction. However, different modalities exhibit substantial discrepancies in semantics, quality, and availability, leading to highly heterogeneous modality combinations and posing significant challenges to achieving consistent and reliable emotion understanding. To address this challenge, we propose the \textbf{Modality-Aware Contrastive and Uncertainty-Regularized (MCUR)} framework, which approaches MER from the perspective of representation consistency, aiming to enable robust emotion prediction across heterogeneous modality combinations.
MCUR incorporates two core components: (1) Modality Combination-Based and Category-Based Contrastive Learning mechanism (MCB-CL), which encourages samples with the same emotion category and the same available modalities to be close in the representation space; and (2) Sample-wise Uncertainty-Guided Regularization (SUGR), which adaptively assigns sample-wise uncertain weights to samples to optimize training. Extensive experiments demonstrate that MCUR consistently outperforms existing methods, achieving average F1 gains of 2.2\% on MOSI, 2.67\% on MOSEI, and 4.37\% on IEMOCAP.
\end{abstract}

\section{Introduction}
\label{sec:intro}
Multimodal emotion recognition (MER) aims to infer the emotional state of an utterance by integrating cues from language, vision, and audio. With the rapid progress of human-computer interaction technologies, MER has become a crucial research direction \cite{lian2025ovmer, fang2025catch,zeng2023multimodal}. However, real-world MER systems still face strong modality heterogeneity, as different modalities exhibit significant discrepancies in:
(1) \emph{Semantics}: text conveys explicit emotional polarity, while audio reflects intensity and rhythm, and visual cues express subtle affective nuances;
(2) \emph{Reliability}: modalities differ in quality, where language is precise but sparse, whereas audio-visual inputs are rich yet easily corrupted by noise;
(3) \emph{Availability}: missing or degraded modalities naturally lead to diverse modality combinations, resulting in drastic input distribution shifts that further exacerbate the above inconsistencies and hinder cross-modal alignment and robust fusion.

\begin{figure}
  \centering
  \includegraphics[width=0.88\linewidth]{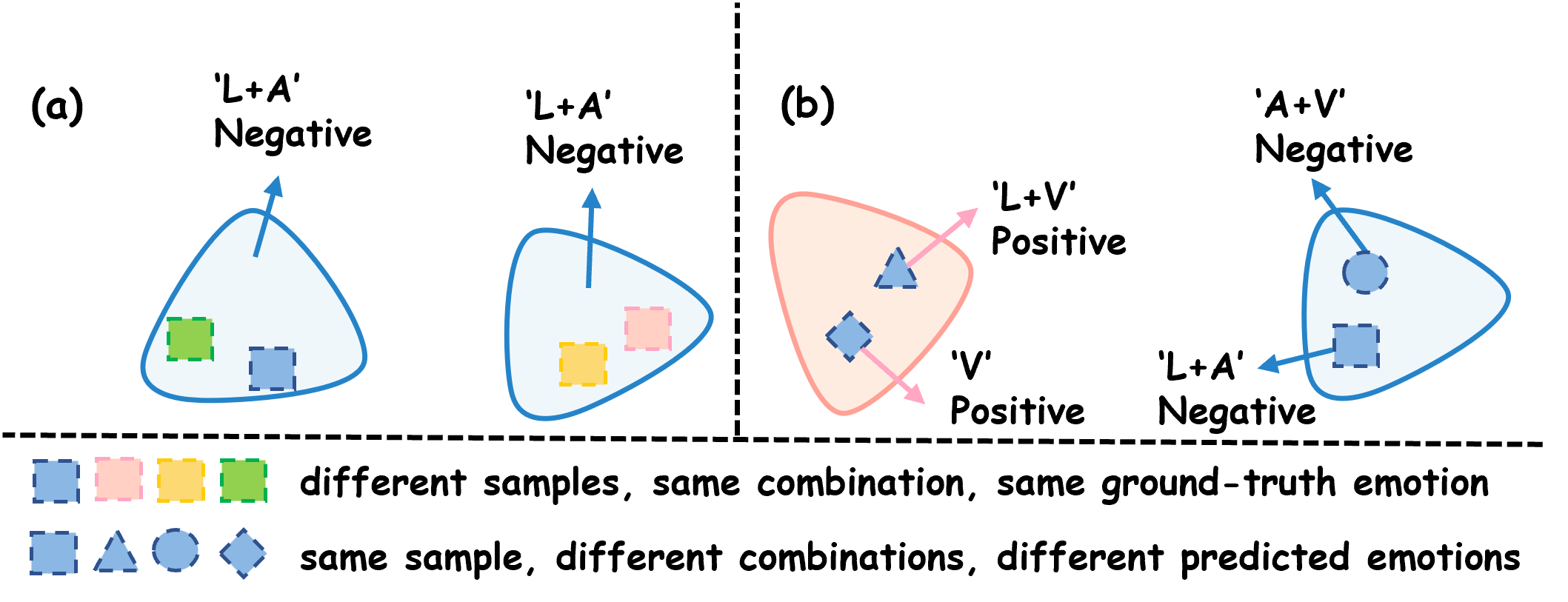}
  \caption{Illustration of the representation inconsistency challenges. (a) Intra-combination inconsistency: Even under the same modality combination (e.g., L+A), samples with the same emotion label (e.g., $\textit{Negative}$) exhibit dispersed representations. (b) Cross-combination inconsistency: A single sample produces contradictory predictions when evaluated with different available modality subsets (e.g., L+V vs.\ A+V).} \label{fig:motivation}
\end{figure}

Previous studies have primarily regarded this as a missing-modality problem, focusing on handling specific missing-modality cases.
For instance, \emph{reconstruction-based methods} aim to infer missing modalities from the available ones (e.g., MPLMM \cite{guo2024multimodal}, DiCMoR \cite{wang2023distribution}, LNLN \cite{zhang2024towards}),
while \emph{knowledge distillation (KD)-based methods} (e.g., CorrKD \cite{li2024correlation}, MMANet \cite{wei2023mmanet}) transfer knowledge from a teacher trained on complete modalities to a student exposed to partial inputs. Although these methods can improve cross-modal complementarity, they primarily focus on recovering missing information or transferring knowledge from complete modalities. However, they seldom impose explicit constraints on the structure of the representation space under different modality combinations. As the available evidence changes, the same emotional semantics may be mapped to different regions of the embedding space, leading to inconsistent representations and suboptimal performance under heterogeneous missing conditions. As illustrated in Figure \ref{fig:motivation}, this inconsistency manifests in two forms: \textbf{(1) Intra-combination inconsistency}, where samples with the same emotion label and the same modality combination are still dispersed in the representation space; and \textbf{(2) Cross-combination inconsistency}, where the same sample yields contradictory predictions under different available modality subsets (e.g., L+V vs. A+V). These observations indicate that neither reconstruction nor knowledge distillation alone can guarantee semantic alignment across modality combinations. Therefore, we shift the focus of missing-modality emotion recognition from reconstructing or distilling missing information to maintaining consistent representations for the same emotional semantics under different modality combinations. 

To tackle these challenges, we propose MCUR, a modality-aware contrastive and uncertainty-regularized framework that explicitly promotes representation consistency and predictive stability across heterogeneous modality settings.
MCUR consists of two key modules: (1) \textbf{Modality Combination-Based and Category-Based Contrastive Learning (MCB-CL)}, which jointly models modality combinations and emotion categories to enforce contrastive constraints across modality subspaces, thereby enhancing both consistency and discriminability; (2) \textbf{Sample-wise Uncertainty-Guided Regularization (SUGR)}, which quantifies the uncertainty discrepancy between teacher and student predictions to dynamically adjust task and distillation weights, achieving adaptive balance across heterogeneous inputs. Our key contributions are summarized as follows:
\begin{itemize}
    \item We approach missing-modality emotion recognition as a representation consistency problem and propose MCUR, a framework for producing robust embeddings under diverse missing-modality configurations.
    \item We introduce MCB-CL, a contrastive learning mechanism that jointly considers modality combinations and label classes to enhance representation consistency and discriminability.
    \item We propose SUGR, an uncertainty-aware regularization strategy that dynamically adjusts training weights based on the uncertainty of each sample.
\end{itemize}

\section{Related Work}\label{related}
\subsection{MER with Missing Modalities}
Multimodal emotion recognition (MER) aims to predict the emotion of a given utterance by integrating signals from language, video, and audio \cite{zhuang2024glomo,fang2025emoe,zhuang2026iie}. Existing approaches to address MER with missing modalities can be broadly categorized into two groups: modality reconstruction-based methods and knowledge distillation (KD)-based methods. Modality reconstruction-based methods estimate the missing modality using the available ones, thereby recovering the original information \cite{wang2023distribution, zhang2024towards, zhuang2026iie,wang2023incomplete, lian2023gcnet,zhuang2026tmdc}. For instance, IMDer \cite{wang2023incomplete} employs a diffusion model to reconstruct missing modalities by comparing the original and reconstructed representations, while DiCMoR \cite{wang2023distribution} models distributions for each modality and category and uses them to guide the reconstruction of missing modalities. MPLMM \cite{guo2024multimodal} pre-trains on complete datasets and uses several prompts to learn information from each modality, which is then applied to reconstruct the missing modalities. LNLN \cite{zhang2024towards}, on the other hand, focuses on the language modality, leveraging it as the dominant source of information in incomplete scenarios. While these methods show promise, they often treat modality reconstruction as an isolated task and struggle to produce robust, unified representations across diverse missing-modality conditions. In contrast, our proposed MCUR, is a KD-based framework that learns directly from fused representations without explicit modality reconstruction. By jointly considering modality combinations and emotion categories, MCUR enhances both consistency and discriminability in learned features, and learns more robust representations across varying missing-modality scenarios.

\subsection{Knowledge Distillation}
Knowledge distillation (KD)-based methods \cite{wei2023mmanet, li2024correlation, hinton2015distilling, tung2019similarity,li2024toward} aim to preserve relationships between samples while reconstructing fused representations. These approaches typically involve a teacher model, trained with full modality data, and a student model, which learns from datasets with missing modalities. The teacher guides the student by providing sample representations \cite{li2024unified}, inter-sample relationships \cite{li2024correlation}, and logits \cite{li2024unified,zhuang2025cmad}, helping improve the student’s performance. For example, MMANet \cite{wei2023mmanet} transfers inter-sample relationships and strengthens weaker modality representations. Although effective, these methods typically overlook consistency within the same emotion category under specific missing-modality patterns, as well as sample-wise differences in prediction uncertainty. In contrast, our proposed MCUR addresses these limitations by leveraging both modality combinations and emotion categories to enhance representation consistency and discriminability, and dynamically adjusting the loss weights based on the prediction uncertainty of each sample. 

\section{Methodology}\label{methods}
\begin{figure*}[t]
\centering
\includegraphics[width=0.88\linewidth]{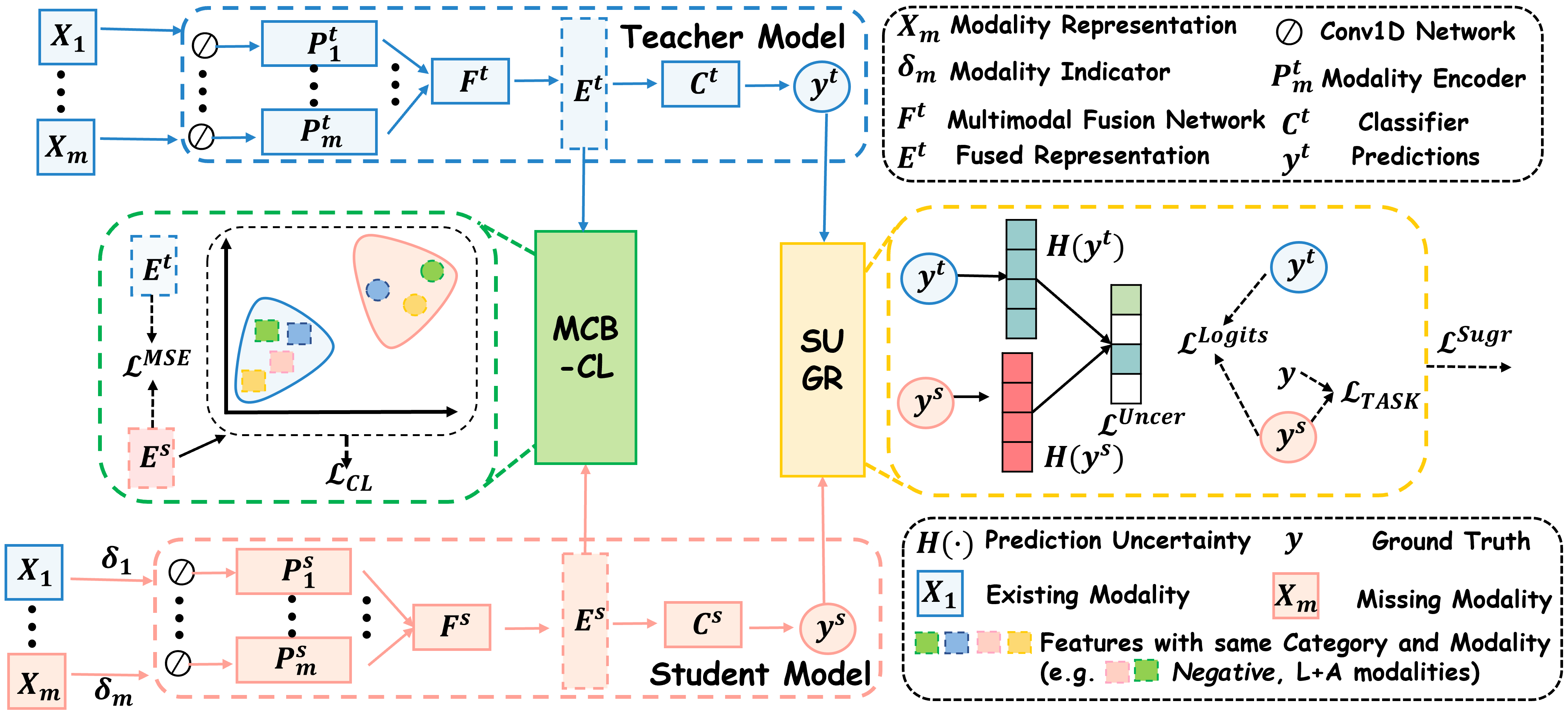} 
\caption{The structure of the MCUR framework. MCUR includes a teacher model, a student model and two key modules: Modality Combination-Based and Category-Based Contrastive Learning (MCB-CL), and Sample-wise Uncertainty-Guided Regularization (SUGR). MCB-CL enforces the samples with same category and same modalities closer in the representation space while SUGR adaptively assigns sample-wise uncertain weights to samples to optimize training.}
\label{fig_framework}
\end{figure*}

The structure of the proposed MCUR is illustrated in Figure \ref{fig_framework}. MCUR involves two modules: Modality Combination-Based and Category-Based Contrastive Learning (MCB-CL) module and Sample-wise Uncertainty-Guided Regularization (SUGR) module.

\subsection{Problem Definition}
In the context of MER with missing modalities, there are $m$ modalities in the dataset $\mathcal{D}=\{X_1,X_2,...,X_m\}$, where $X_p\in\mathbb{R}^{t_p\times d_p}$ denotes the modality representations and $p\in\{1,2,...,m\}$ denotes the modality. Here $d_p$ and $t_p$ represent the dimensionality and sequence length for each modality, respectively. The goal is to train a robust model that can handle samples with various missing modalities simultaneously. Specifically, for modality $p$, we define $\delta_p\in\mathbb{R}^{N}$ to indicate the presence or absence of the modality. Here $N$ is the size of the dataset. Each element $\delta_{p,i}$ in $\delta_p$ is either 0 (missing modality) or 1 (available modality), with the condition that $\delta_{1,i}+\delta_{2,i}+...+\delta_{m,i}>0$, meaning that each sample must have at least one modality available.

\subsection{MCUR Overview}
MCUR follows the standard KD paradigm \cite{wei2023mmanet, li2024correlation, hinton2015distilling}, employing a teacher-student architecture, as shown in Figure \ref{fig_framework}. The teacher model is pre-trained using full modalities and the student model is trained with limited modalities. And the parameters in the teacher model are frozen when the student model is training. Given multimodal input $x = \{X_1, X_2, ..., X_m\}$, we randomly drop certain modalities (e.g., $X_j$, $X_k$) in the student model by setting $\delta_j = \delta_k = 0$ to simulate real-world missing scenarios. All modalities are first passed through a 1D convolutional layer to capture temporal patterns and standardize their shape to $X_i \in \mathbb{R}^{T \times D}$, where $T$ and $D$ are fixed hyper-parameters. Each modality is then encoded via a modality-specific Perceiver encoder $P^w_j$ \cite{jaegle2021perceiver, liang2022high, zhang2023learning,zhuang2025hyper}, where $j \in \{1, ..., m\}$ and $w \in \{s, t\}$ denotes student or teacher. Encoded features are fused by a transformer encoder \cite{vaswani2017attention} followed by an MLP \cite{hazarika2020misa} and a Variational Information Bottleneck (VIB) \cite{alemi2022deep,gao2024embracing,kingma2013auto, tishby2000information,xiao2024neuro}, forming the fusion module $F^w$, which outputs the fused embedding $E^w \in \mathbb{R}^{D}$. These embeddings from both models are further aligned through the MCB-CL module to ensure representation consistency. Finally, the classifier $C^w$ predicts logits $y^w \in \mathbb{R}^k$, which, along with ground truth $y \in \mathbb{R}^k$, are used by the SUGR module to dynamically weight supervision based on prediction uncertainty. Implementation details of teacher model, $P^w_j$ and $F^w$ are provided in the Appendix \ref{sec:backbone}.

\subsection{MCB-CL}\label{sec:CL}
In this section, we present the Modality Combination-Based and Category-Based Contrastive Learning (MCB-CL) mechanism. Existing methods for handling missing modalities often focus on reconstructing the representations to improve predictions, typically expressed as $p(k|x_i)$, where $x_i$ represents the $i^{th}$ sample's representation and $k$ is its corresponding category \cite{huan2023unimf}. Some approaches also leverage CL \cite{li2024correlation, zhuang2025cmad}, considering representations with the same label as positive pairs. While these methods show promising results, they overlook intra-class consistency under the same missing modality patterns. Specifically, these methods fail to consider the modality combinations ($c_i$), which can provide valuable information, especially when certain modality combinations under-perform. 

To address these limitations, MCB-CL integrates both modality combinations and emotion classes into the CL framework. Specifically, MCB-CL improves $p(k|x_i)$ by incorporating modality combinations $c_i$ alongside it. Using Bayes' theorem and the law of total probability, we rewrite $p(k|x_i)$ as:
\begin{equation}
p(k|x_i)=p(c_i|x_i)\frac{p(k|c_i,x_i)}{p(c_i|x_i,k)}.
\end{equation}
Inspired by existing works \cite{peng2024knowledge, shen2021you, tsai2020mice}, we also apply CL to parametrize each term. 

\textbf{Parametrization of $p(c_i|x_i)$.} First, for $p(c_i|x_i)$, which identifies samples within the batch that share the same modality combination as the current sample's representation, we use supervised multi-instance CL as:
\begin{equation}
p(c_i|x_i)\triangleq\frac{\sum_{j\in\mathcal{M}_i}exp(x_ix_j^\top/\tau)}{\sum_{l=1}^Bexp(x_ix_l^\top/\tau)}.\label{eq6}
\end{equation}
Here $B$ is the batch size, and $\tau$ is temperature hyper-parameter controlling the concentration level. $\mathcal{M}_i=\{j|x_j\in E^s, c_j=c_i, j\le B\}$ represents the indices of samples in the current batch with the same modality combination as $x_i$. 

\textbf{Parametrization of $p(k|c_i,x_i)$.} Next, for $p(k|c_i,x_i)$, which focuses on identifying samples with the same category within a given modality combination, we also apply multi-instance supervised CL. To avoid introducing additional variables, we use $j$ and $l$ for samples during the search:
\begin{equation}
p(k|c_i,x_i)\triangleq\frac{\sum_{j\in\mathcal{N}_i}exp(x_ix_j^\top/\tau)}{\sum_{l\in\mathcal{M}_i}exp(x_ix_l^\top/\tau)}.
\end{equation}
Here, $\mathcal{N}_i=\{j|x_j\in E^s, c_j=c_i\&k_j=k, j\le B\}$ represents the indices of samples with both the same modality combination $c_i$ and the same category $k$ as $x_i$. $\mathcal{M}_i$ is the same as in Equation \ref{eq6}, but $j$ is replaced by $l$ here.

\textbf{Parametrization of $p(c_i|x_i,k)$.} Finally, for $p(c_i|x_i,k)$, which aims to find modality combinations corresponding to a specific category, we again apply supervised CL, using $j$ and $l$ for samples during the search:
\begin{equation}
p(c_i|x_i,k)\triangleq\frac{\sum_{j\in\mathcal{N}_i}exp(x_ix_j^\top/\tau)}{\sum_{l\in\mathcal{S}_i}exp(x_ix_l^\top/\tau)}.
\end{equation}
Here, $\mathcal{S}_i=\{l|x_l\in E^s, k_l=k, l<B\}$ is the indices of samples with the same category $k$ as $x_i$.

While the Bayesian identity provides a theoretical foundation for joint probability decomposition, the implicit coupling between these terms in a discrete contrastive setting may limit the model's ability to fine-tune the importance of different alignment goals. To provide more explicit control over the modality-aware feature alignment and enhance optimization stability, we decouple the components of the Bayesian identity by introducing two hyperparameters: $\mu_1$ and $\mu_2$. Consequently, the final MCB-CL loss for student representations $E^s$ is formulated as:
\begin{equation}
\mathcal{L}_{CL}^{s} = -\mathbb{E}_i \left[ \log p(c_i|E_i^s) + \mu_1 \log p(k|c_i,E_i^s) - \mu_2 \log p(c_i|E_i^s,k) \right].
\end{equation}

Consistent with existing KD methods \cite{li2024correlation, zhuang2025cmad}, we also preserve the consistency between the teacher's and student’s representations, which is achieved by calculating the representation difference through:
\begin{equation}
\mathcal{L}^{MSE}_i=||E^s_i-E^t_i||^2.
\end{equation}

\subsection{SUGR}\label{sec:SUGAR}
In this section, we introduce the Sample-wise Uncertainty-Guided Regularization (SUGR) module to address the overlooked cross-combination inconsistency in prediction uncertainty. Unlike prior work \cite{wei2023mmanet}, which treats all missing-modality cases uniformly, SUGR adaptively emphasizes samples where the absence of certain modalities leads to greater uncertainty and assigns higher weights.

To quantify the prediction uncertainty, we adopt different measures for classification and regression tasks. For classification, we compute the entropy of the predicted probability distribution \cite{wei2023mmanet} :
\begin{equation}
H(y^w_i) = -\sum\sigma(y^w_i)\log(\sigma(y^w_i)),
\end{equation}
where $\sigma(y^w_i)$ is the softmax probability output of the model. For regression, the Mean Squared Error (MSE) between the predicted value and the ground truth is applied:
\begin{equation}
H(y^w_i) = |y_i - y^w_i|^2.
\end{equation}
The absolute uncertainty difference between the teacher and student uncertainty is then computed to measure the increased uncertainty:
\begin{equation}
\mathcal{L}^{Uncer}_i = |H(y^t_i) - H(y^s_i)|,
\end{equation}
where a larger discrepancy implies higher uncertainty and leads to a greater penalty during optimization.

To ensure reliable prediction, logits distillation is also adapted as in prior work \cite{li2024correlation, li2024unified, zhuang2025cmad}. For regression tasks, the MSE between teacher and student predictions is adapted:
\begin{equation}
\mathcal{L}^{Logits}_i = |y^s_i - y^t_i|^2.
\end{equation}
For classification tasks, we use Decoupled Knowledge Distillation (DKD) \cite{zhao2022decoupled}, which separately distills both target and non-target class probabilities:
\begin{equation}
DKD(y^s, y^t) = \alpha KL(b^t || b^s) + (1 - \alpha) KL(\hat{p}^t || \hat{p}^s),
\end{equation}
where $b^t$ and $b^s$ are the predicted probabilities of the target class, and $\hat{p}^t$ and $\hat{p}^s$ correspond to non-target classes. $\alpha$ is a hyper-parameter balancing the two components.

Additionally, the task-specific loss $\mathcal{L}_{TASK}$ is adapted for downstream tasks. For regression, we use the Mean Absolute Error:
\begin{equation}
\mathcal{L}_{TASK,i}^s = |y_i - y^s_i|.\label{TASK1}
\end{equation}
For classification, we use standard cross-entropy loss:
\begin{equation}
\mathcal{L}_{TASK,i}^s = CrossEntropy(y_i, y^s_i).\label{TASK2}
\end{equation}

Finally, all above loss terms are combined into the overall SUGR loss. The uncertainty difference modulates the influence of both the task-specific loss and the logits distillation loss:
\begin{equation}
\mathcal{L}^{Sugr}_{i} = \mathcal{L}^{Uncer}_{i} \cdot (\mathcal{L}^{s}_{TASK,i} + \mathcal{L}^{Logits}_i).
\end{equation}

\textbf{Training Objectives.} The overall training loss for MCUR framework is obtained through:
\begin{equation}
\mathcal{L}_{all}=\gamma\mathcal{L}_{CL}^{S}+\mathcal{L}^{MSE}+\zeta \mathcal{L}^{Sugr},
\end{equation}
where $\gamma$, and $\zeta$ are hyper-parameters.

\section{Experiments}\label{experi}

\begin{table*}[ht]
\centering
\caption{Performance comparison in datasets. `ACC/F1' is reported. Float values in `Avail' are MR.}\label{EXPERI_all_fix}
\begin{tabular}{ccccccc}
\hline
Avail & CorrKD & MPLMM & IMDer & LNLN & MMANet & MCUR \\
\hline
\hline
\multicolumn{7}{c}{Results on MOSEI} \\
\hline
L & 85.66/85.64 & 84.87/84.94 & 85.31/85.01 & 84.98/85.03 & 85.50/85.34 & \textbf{85.97/85.88} \\
V & 63.65/48.51 & 61.39/53.49 & 62.33/57.04 & 62.47/51.44 & 62.96/58.75 & \textbf{65.00/64.59} \\
A & 62.85/48.51 & 62.80/48.49 & 62.85/48.51 & 62.85/48.51 & 63.29/51.54 & \textbf{63.48/60.35} \\
L,V & 86.08/86.02 & 85.09/85.09 & 85.31/85.31 & 85.80/85.80 & 85.44/85.27 & \textbf{86.13/86.07} \\
L,A & 85.69/85.68 & 84.84/84.89 & 85.33/85.00 & 85.53/85.54 & 85.55/85.43 & \textbf{85.97/85.88} \\
A,V & 64.00/55.83 & 62.66/50.85 & 61.97/53.23 & 62.82/49.42 & 64.34/61.41 & \textbf{64.36/64.42} \\
L,A,V & 85.99/85.97 & 85.20/85.13 & 85.86/85.84 & 86.02/85.98 & 84.98/84.99 & \textbf{86.10/86.04} \\
0.1 & 83.16/82.96 & 83.02/82.78 & \textbf{83.82/83.64} & 83.71/83.48 & 81.89/81.77 & 83.46/83.39 \\
0.2 & \textbf{81.98/81.56} & 80.49/79.97 & 80.96/80.37 & 81.40/80.88 & 80.21/79.91 & 81.26/81.17 \\
0.3 & 79.06/78.25 & 79.28/78.45 & 79.00/78.34 & 79.00/78.06 & 78.23/77.56 & \textbf{79.88/79.76} \\
0.4 & 77.02/75.72 & 75.98/74.36 & 73.20/70.89 & 77.41/76.03 & 76.97/75.99 & \textbf{77.55/77.32} \\
0.5 & 75.18/73.54 & 73.45/71.21 & 69.90/66.38 & 73.69/71.14 & 73.91/72.16 & \textbf{75.89/75.64} \\
0.6 & 70.64/67.56 & 71.79/68.84 & 72.15/69.27 & 71.41/67.79 & 72.29/69.72 & \textbf{73.50/72.87} \\
0.7 & 70.61/67.51 & 69.21/65.52 & 65.58/59.33 & 70.53/66.24 & 70.83/67.94 & \textbf{72.59/71.82} \\
Avg. & 76.54/73.88 & 75.72/72.43 & 75.26/72.01 & 76.26/72.52 & 76.17/74.13 & \textbf{77.23/76.80} \\
\hline
\hline
\multicolumn{7}{c}{Results on IEMOCAP} \\
\hline
L & 75.37/72.46 & 75.83/70.03 & 75.03/64.72 & 76.49/72.55 & 73.85/71.86 & \textbf{79.58/77.89} \\
V & 74.57/66.20 & 71.91/64.77 & 75.00/65.07 & 74.60/67.85 & 75.08/67.38 & \textbf{75.40/70.23} \\
A & 77.05/73.25 & 76.79/68.62 & 74.65/65.50 & 76.55/68.74 & 75.00/64.66 & \textbf{79.42/76.84} \\
L,V & 75.99/73.34 & 75.69/70.47 & 75.32/65.90 & 76.57/73.42 & 74.01/71.75 & \textbf{79.64/78.10} \\
L,A & 78.68/76.00 & 76.23/70.58 & 75.35/66.10 & 78.33/75.23 & 73.80/71.77 & \textbf{81.50/80.23} \\
A,V & 76.68/73.32 & 72.84/65.54 & 75.64/67.65 & 76.55/71.50 & 74.92/67.26 & \textbf{79.48/77.47} \\
L,A,V & 78.89/76.31 & 76.01/70.58 & 75.72/67.85 & 78.46/75.87 & 73.93/71.66 & \textbf{81.74/80.72} \\
0.1 & 77.99/75.29 & 75.64/70.02 & 75.51/67.32 & 78.09/75.30 & 74.49/72.01 & \textbf{81.69/80.51} \\
0.2 & 77.83/75.20 & 75.67/69.88 & 75.19/66.88 & 78.28/75.13 & 74.28/71.57 & \textbf{80.44/79.09} \\
0.3 & 77.56/74.54 & 75.27/69.22 & 75.16/66.56 & 77.88/74.45 & 74.47/71.04 & \textbf{80.49/79.01} \\
0.4 & 76.84/73.72 & 74.39/68.46 & 75.40/66.40 & 76.44/72.52 & 74.87/71.42 & \textbf{80.68/78.98} \\
0.5 & 76.20/72.69 & 75.05/68.70 & 74.95/65.68 & 76.97/72.60 & 74.57/70.32 & \textbf{78.81/76.56} \\
0.6 & 75.91/71.92 & 75.00/68.46 & 75.13/65.46 & 76.23/70.80 & 74.25/69.33 & \textbf{78.60/76.12} \\
0.7 & 75.99/71.66 & 74.44/67.58 & 74.84/65.04 & 75.21/69.63 & 75.00/69.78 & \textbf{78.14/75.42} \\
Avg. & 76.82/73.28 & 75.05/68.78 & 75.21/66.15 & 76.90/72.54 & 74.47/70.13 & \textbf{79.69/77.65} \\
\hline
\end{tabular}
\end{table*}

\subsection{Datasets and Evaluation Metrics}
In line with prior work \cite{guo2024multimodal, li2024correlation}, we evaluate the MCUR framework on three widely used MER datasets: MOSI \cite{zadeh2016mosi}, MOSEI \cite{zadeh2018multimodal}, and IEMOCAP \cite{busso2008iemocap}. For the MOSI and MOSEI datasets, we use accuracy (ACC) and F1 score computed for `positive/negative' classification results as evaluation metrics. For IEMOCAP dataset, we calculate the weighted accuracy and F1 score of four emotion categories.

\subsection{Implementation Details}
We evaluate model performance under two protocols: fixed missing and random missing. In the fixed missing protocol, we test each model with one specific modality missing. For instance, in Table \ref{EXPERI_all_fix}, the label `L' indicates that only the language modality is available. In the random missing protocol, modalities are randomly dropped for each sample, as in prior work \cite{lian2023gcnet, wang2023incomplete, wang2023distribution}. To quantify the extent of missing modalities, we used the missing rate (MR), defined as: $MR = 1-\frac{\sum_{i=1}^{N}a_i}{N\times m}$. Here $a_i$ represents the number of available modalities in the $i^{th}$ sample, $N$ is the number of samples and $m$ is the number of modalities in each sample. Given that each sample must contain at least one modality, we ensure that $a_i>0$ and $MR\le \frac{m-1}{m}$. The MR values are set between 0.1 and 0.7, with 0.7 approximating the maximum missing rate $\frac{2}{3}$. More implementation details are shown in Appendix \ref{sec:addi_imple}. 

\subsection{Comparison with State-of-the-art Methods} 
We compare MCUR with several SOTA methods, including both modality reconstruction-based models (MPLMM \cite{guo2024multimodal}, IMDer \cite{wang2023incomplete}, and LNLN \cite{zhang2024towards}) and KD-based models (CorrKD \cite{li2024correlation}, MMANet \cite{wei2023mmanet}). To ensure fairness, we use official implementations and apply the same training settings across all models. Each model is evaluated under 14 missing modality scenarios—seven fixed and seven random. Additional results on MOSI dataset are reported in the Appendix\ref{sec:additional_results}.

Results are presented in Table \ref{EXPERI_all_fix}, and `Avg' denotes the average performance over all scenarios, with the best scores highlighted in bold. MCUR consistently outperforms all baselines across most scenarios and datasets. On average, it surpasses others by 0.69/2.67 and 2.79/4.37 in ACC/F1 across the 14 test cases, demonstrating strong representation consistency. Interestingly, KD models generally outperform reconstruction ones. This may be because reconstruction methods focus on recovering individual modalities, which can accumulate errors during fusion. In contrast, KD models directly guide the student to approximate the fused representations, reducing such error amplification. We also observe that for most baseline models, accuracy scores are notably higher than F1 scores, suggesting a tendency to overfit to frequent emotions. In contrast, MCUR yields more balanced accuracy and F1 results, indicating better representation consistency and more stable predictions.

\begin{table}[t]
\centering
\caption{Ablation results of MCUR components. Averaged performance is reported.}\label{abla_datasets}
\begin{tabular}{cccc|cccc}
\hline
& MOSI & MOSEI & IEMOCAP & \multicolumn{2}{c}{MOSEI} & \multicolumn{2}{c}{IEMOCAP} \\
\hline
Models & ACC/F1 & ACC/F1 & ACC/F1 & Brier & NLL & Brier & NLL \\
\hline
Full & \textbf{74.5/74.0} & \textbf{77.2/76.8} & \textbf{79.7/77.7} & \textbf{0.179} & \textbf{0.541} & \textbf{0.159} & \textbf{0.492} \\
w/o $\mathcal{L}_{CL}^{s}$ & 71.6/70.5 & 75.4/75.5 & 77.9/75.1 & 0.183 & 0.547 & 0.174 & 0.530 \\
w/o $\mathcal{L}^{Uncer}$ & 70.6/70.5 & 75.9/74.4 & 76.6/75.2 & 0.182 & 0.545 & 0.186 & 0.556 \\
w/o $\mathcal{L}^{Logits}$ & 71.5/71.4 & 76.3/75.8 & 78.8/76.5 & 0.182 & 0.545 & 0.175 & 0.532 \\
w/o $\mathcal{L}^{MSE}$ & 72.7/71.8 & 76.3/75.8 & 78.4/75.5 & 0.181 & 0.543 & 0.173 & 0.525 \\
\hline
\end{tabular}
\end{table}

\subsection{Ablation Studies}
To evaluate the contribution of each component in MCUR, we perform ablation studies, with results shown in Table \ref{abla_datasets}. Specifically, we assess the impact of four loss terms: the contrastive learning loss ($\mathcal{L}_{CL}^{s}$), the uncertainty term ($\mathcal{L}^{Uncer}$), the logits distillation loss ($\mathcal{L}^{Logits}$), and the MSE loss ($\mathcal{L}^{MSE}$), across 14 missing modality settings.

Removing any of these components leads to an average performance drop of over 0.88 in ACC and 1.0 in F1. The most critical components are the $\mathcal{L}_{CL}^{s}$ and $\mathcal{L}^{Uncer}$. In particular, excluding the uncertainty loss causes substantial performance declines: 3.89/3.47 on MOSI, 1.29/2.38 on MOSEI, and 3.12/2.5 on IEMOCAP, respectively. We also find that the impact of each component varies by datasets. In MOSI and MOSEI, removing the MSE loss leads to the smallest degradation, suggesting that representation alignment is less critical. In IEMOCAP, the smallest drop comes from removing the logits distillation loss.

\begin{figure*}[ht]
  \centering
  \includegraphics[width=0.24\linewidth]{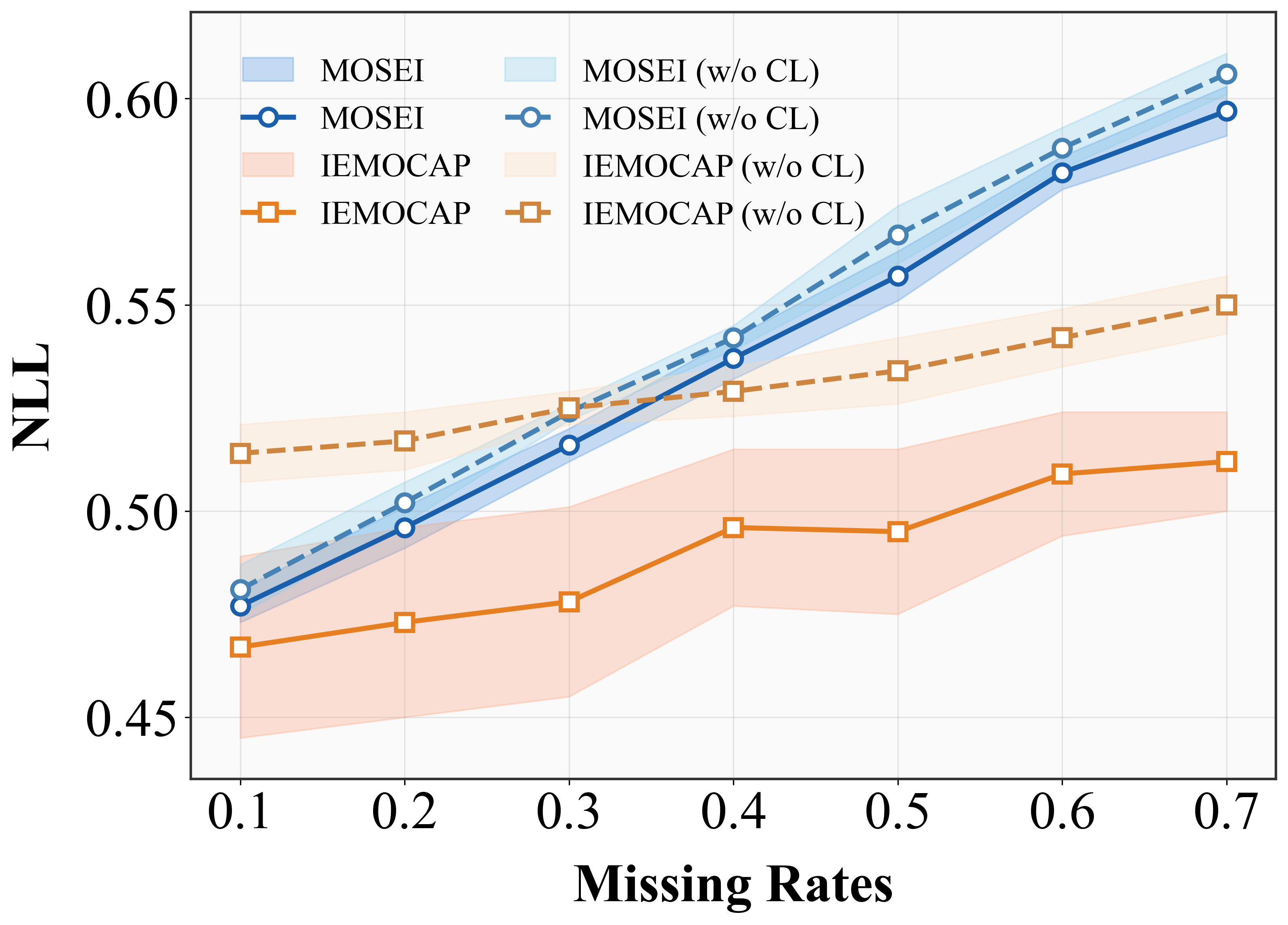}\label{fig_NLL_MR}
  \includegraphics[width=0.24\linewidth]{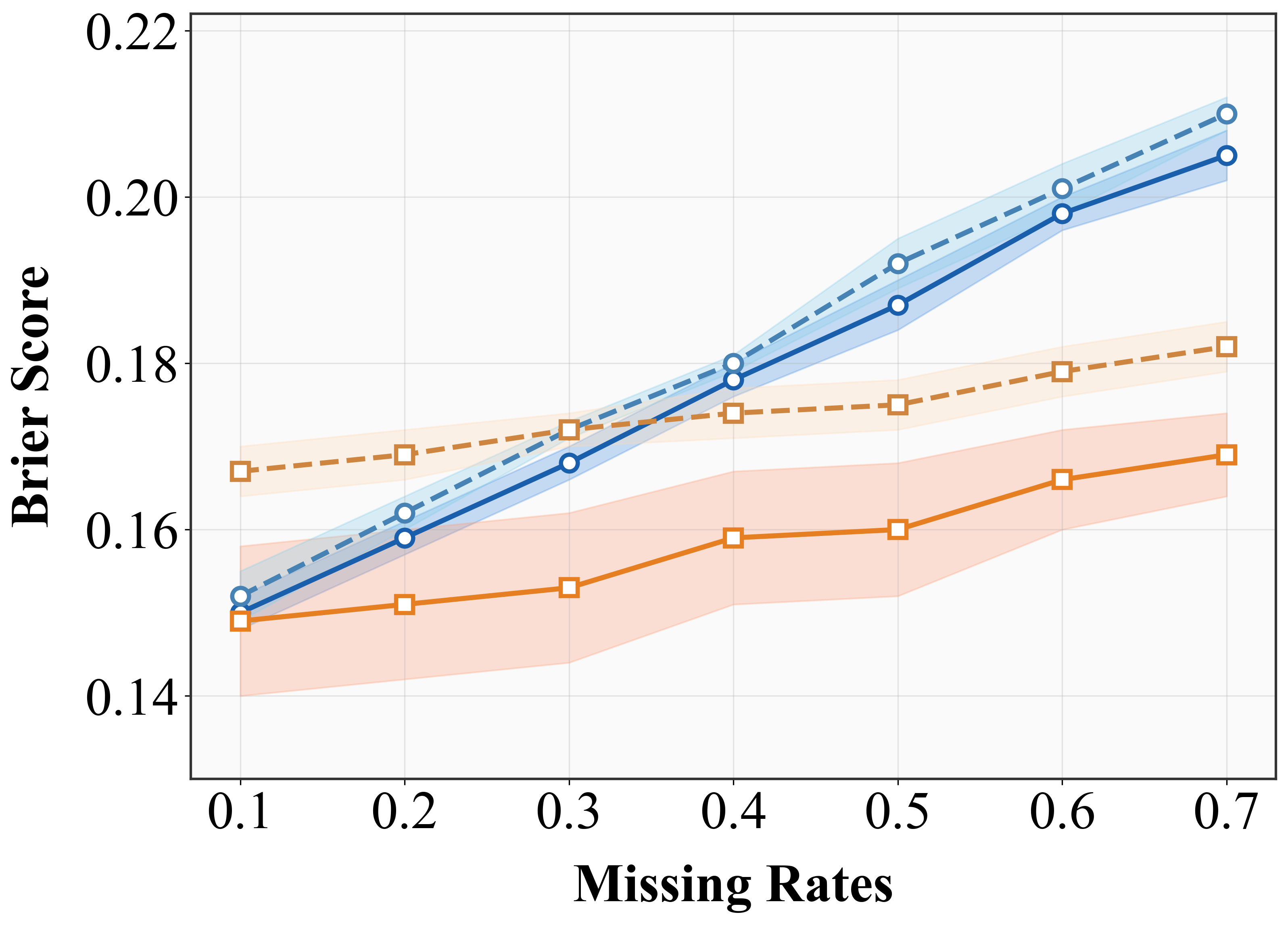}\label{fig_Brier_MR}
  \includegraphics[width=0.24\linewidth]{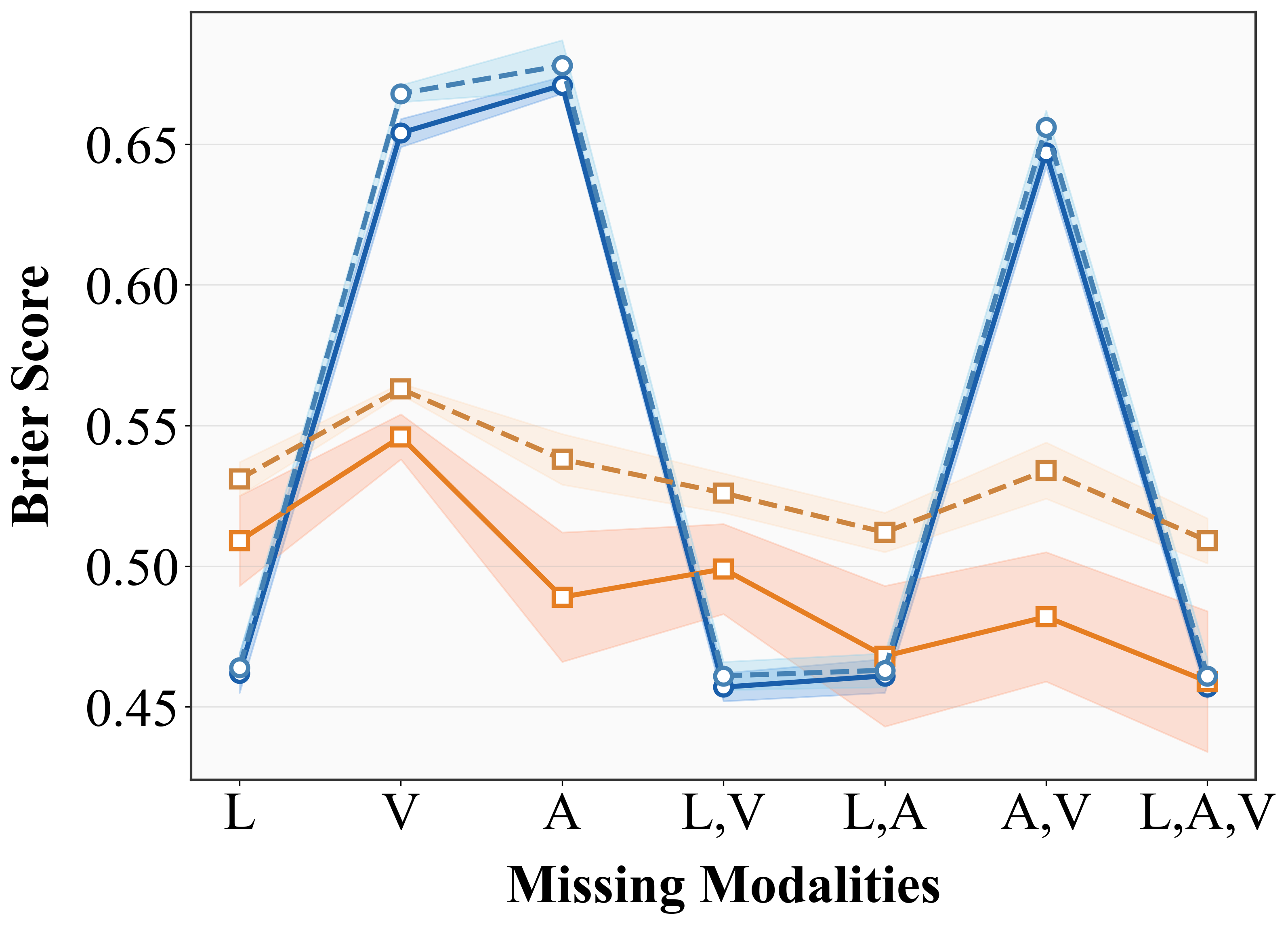}\label{fig_NLL_MM}
  \includegraphics[width=0.24\linewidth]{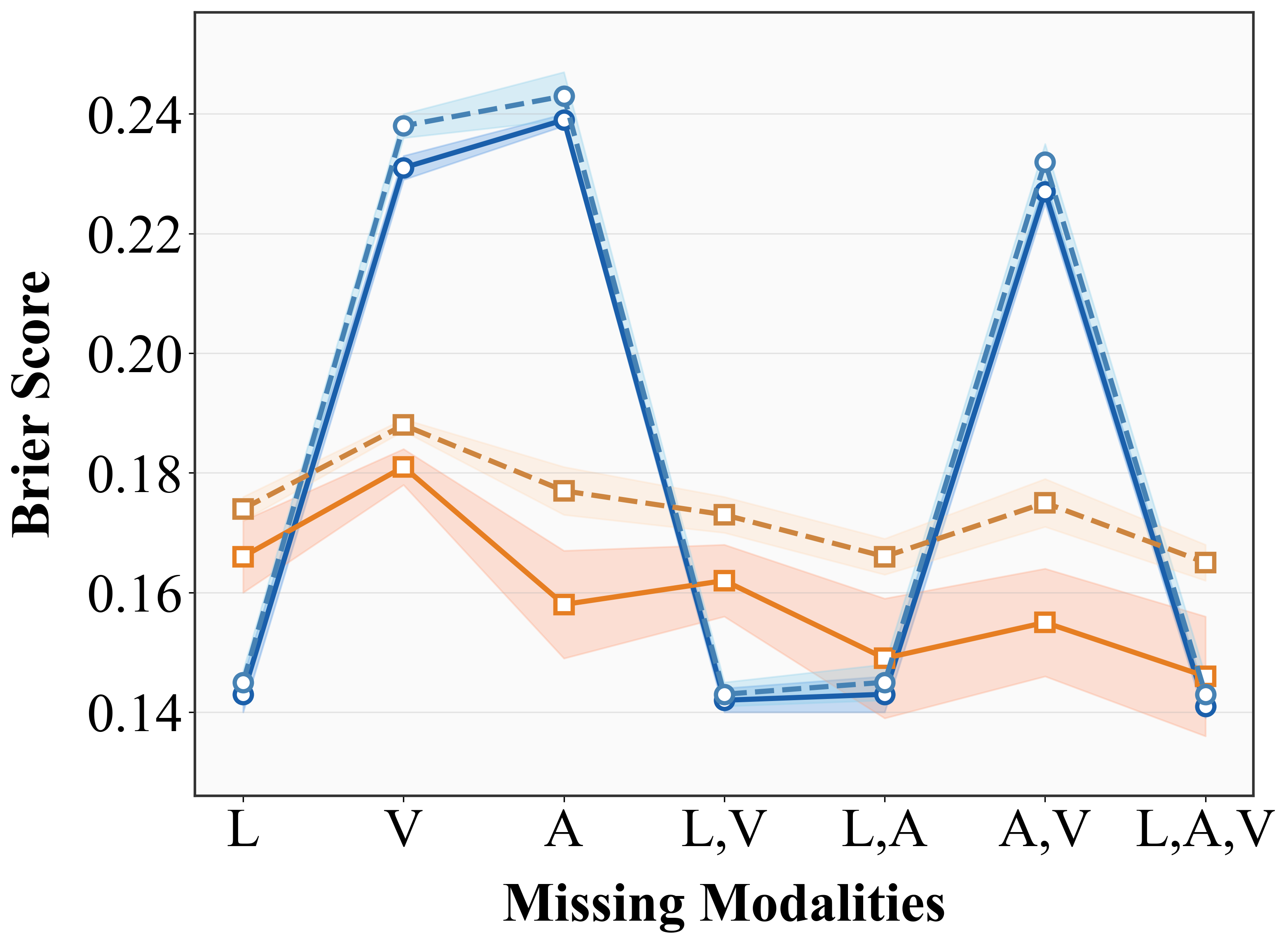}\label{fig_Brier_MM}
 \caption{Uncertainty estimation under different missing modality conditions. (a) NLL under different MRs; (b) Brier under different MRs; (c) NLL under different modalitie; (d) Brier under different modalities. The Brier Score and NLL are used as evaluation metrics. Solid lines represent the full MCUR model, while dashed lines indicate the variant without contrastive learning. Orange curves show results on the IEMOCAP dataset, and blue curves correspond to the MOSEI dataset.}\label{fig_UNCER}
\end{figure*}

\subsection{Quantitative Analysis}

\subsubsection{Effects of MCB-CL}
To better understand the necessity of integrating combination into CL and to examine whether it mitigates intra-combination inconsistency, we conduct ablation studies on the MOSEI and IEMOCAP datasets. The prediction uncertainty is used as the measurement of inconsistency and the visualization of the uncertainty under different missing-modality conditions, with and without MCB-CL, is shown in Figure \ref{fig_UNCER}. Specifically, we adopt two standard uncertainty metrics: the Brier Score, which measures the calibration of probabilistic predictions, and the Negative Log-Likelihood (NLL), which quantifies the divergence between predicted probabilities and the ground truth.

Under the random missing protocol, uncertainty increases monotonically with the MR, regardless of whether MCB-CL is applied. On MOSEI, uncertainty grows rapidly but remains stable across runs, while on IEMOCAP, it increases more slowly yet exhibits larger variance. This difference stems from the task characteristics: MOSEI is a regression-based dataset with a thresholded decision boundary, whereas IEMOCAP involves discrete emotion categories that introduce greater intrinsic uncertainty. When MCB-CL is removed (dashed lines in the Figure \ref{fig_UNCER}), both datasets show a clear increase in uncertainty, with IEMOCAP displaying a larger rise than MOSEI. This demonstrates that MCB-CL enhances the robustness and consistency of representations. Interestingly, on IEMOCAP, incorporating MCB-CL increases the variance of uncertainty, but its maximum uncertainty remains lower than the minimum when MCB-CL is absent.

Under the fixed missing protocol, the overall trend remains similar, though the uncertainty distribution varies depending on which modality is missing. Consistent with the random setting, MOSEI exhibits smaller fluctuations, while IEMOCAP shows larger ones. However, the absolute uncertainty differs more markedly across missing-modality types. On MOSEI, maintaining the `A', `V', or `A,V' modalities results in significantly higher uncertainty, reflecting the dominant predictive role of the `L' modality \cite{zhuang2025cmad}. In contrast, on IEMOCAP, where different modalities contribute more evenly, the uncertainty variation across missing-modality conditions is less pronounced. Across both datasets, removing MCB-CL consistently increases uncertainty and amplifies intra-combination inconsistency. Notably, on MOSEI, MCB-CL leads to a larger reduction in uncertainty for modality combinations that originally exhibit higher inconsistency, such as `V', `A', and `A,V'. This suggests that MCB-CL is particularly effective when modality representations are less aligned.

\subsubsection{Uncertainty Analysis}
To further investigate cross-combination inconsistency, we analyze the predictive uncertainty of different MCUR variants. As before, we use the Brier Score and NLL as evaluation metrics, and the results are summarized in Table \ref{abla_datasets}. For a more comprehensive comparison, we report the average uncertainty across all 14 missing-modality settings, averaged over five random runs. 

We observe that the full MCUR model consistently exhibits the lowest uncertainty, while removing any component leads to a noticeable increase in uncertainty. On MOSEI, excluding the $\mathcal{L}_{CL}^{s}$ causes the largest rise in uncertainty, indicating that MCB-CL effectively strengthens representation alignment and improves stability, which aligns with the trends shown in Figure \ref{fig_UNCER}. Removing the uncertainty loss $\mathcal{L}^{Uncer}$ produces the second-largest degradation. On IEMOCAP, omitting $\mathcal{L}^{Uncer}$ results in the most significant increase in uncertainty, suggesting that uncertainty-aware weighting provides critical guidance. This may lies in IEMOCAP’s multi-class emotion recognition setting, which naturally contains higher ambiguity.

\subsubsection{Effects of Different CL Formulations}
To investigate the efficacy of the proposed MCB-CL mechanism, we compare it against four contrastive learning variants: (1) \textbf{Traditional CL} \cite{radford2021learning}, which treats each sample as a unique instance and all others as negatives, ignoring label information;
(2) \textbf{SupCon-CL} \cite{khosla2020supervised}, which considers all samples within the same emotion category as positives;
(3) \textbf{Vanilla MCB-CL (Ours)}, which defines positives as samples sharing both the same emotion category and the identical modality-missing pattern. The average performance across 14 missing modality scenarios is summarized in Table \ref{Abla_cl}; and (4) \textbf{1-instance MCB-CL}, a simplified version that selects only a single positive sample per anchor. As shown in Table \ref{Abla_cl}, Traditional CL yields the lowest performance, likely because it neglects semantic labels and inadvertently penalizes semantically related samples as negatives. While the 1-instance variant and SupCon-CL show notable improvements by incorporating label-level supervision, they both fall short of our proposed Vanilla MCB-CL. Specifically, our method achieves the superior results, underscoring that simultaneously accounting for category consistency and modality-specific patterns provides a more fine-grained and discriminative supervision signal for robust multimodal representation learning.

\begin{table}[t]
\centering
\caption{Ablation of CL formulations in MCUR. Averaged `ACC/F1' is reported.}\label{Abla_cl}
\begin{tabular}{ccccc}
\hline
Dataset & Vanilla & Traditional & 1-instance & SupCon \\
\hline
MOSI & \textbf{74.46/73.99} & 71.47/70.72 & 71.49/71.05 & 73.72/68.28 \\
MOSEI & \textbf{77.23/76.80} & 74.74/74.39 & 75.38/75.27 & 75.99/75.85 \\
\hline
\end{tabular}
\end{table}

\subsubsection{Generalizability of the MCUR Components}\label{sec_4_5_4}
To evaluate the generalizability of MCUR’s core components, we integrated MCB-CL and SUGR into two representative architectures, CorrKD and MMANet. Evaluation across 14 missing modality configurations in Table \ref{tbl_generation} reveals consistent performance gains regardless of the base model. Specifically, incorporating either module into CorrKD yielded improvements of up to 1.53\% on IEMOCAP. These enhancements were even more pronounced for MMANet, where MCB-CL and SUGR boosted performance by up to 4.14\% and 4.54\% on IEMOCAP, respectively, while also providing steady gains on MOSEI. These results underscore the "plug-and-play" capability of our modules and their effectiveness in bolstering model robustness against modality incompleteness.

\begin{table}{}
\centering
\caption{Generalizability of the MCUR components on KD baselines. Averaged `ACC/F1' is reported.}\label{tbl_generation}
\begin{tabular}{c|ccc|ccc}
\hline
& \multicolumn{3}{c|}{MOSEI} & \multicolumn{3}{c}{IEMOCAP} \\
\hline
Models & Vanilla & +MCB-CL & +SUGR & Vanilla & +MCB-CL & +SUGR \\
\hline
CorrKD & 76.5/73.9 & 76.6/74.9 & \textbf{76.8/75.3} & 76.8/73.3 & \textbf{78.1/74.8} & 77.9/74.8 \\
MMANet & 76.2/74.1 & \textbf{76.5/75.3} & 76.3/75.1 & 74.5/70.1 & \textbf{78.1}/74.3 & 78.1/\textbf{74.7} \\
\hline
\end{tabular}
\end{table}

Additional results, including analysis of robustness in real-world scenarios (noisy and missing modalities), computation overhead, cross-dataset transfer, hyper-parameters, teacher bias on language modality, training loss convergence and representation visualization are provided in the Appendix\ref{sec:additional_results}.

\section{Conclusion}\label{conclusion}
In this paper, we presented MCUR, a unified framework that tackles MER from the perspective of representation consistency. MCUR integrates a modality–combination and category–aware contrastive learning (MCB-CL) mechanism to promote discriminative and consistent representations across heterogeneous modality configurations. In addition, the proposed SUGR module dynamically calibrates the training process by weighting samples based on their predictive uncertainty, fostering more stable and reliable learning. Extensive experiments on three benchmark datasets demonstrate the effectiveness and robustness of MCUR under diverse modality conditions. Future work will extend MCUR to handle noisy or degraded modalities in more realistic multimodal environments.

\bibliographystyle{IEEEtran}
\bibliography{neurips_2026}


\appendix
\section{Appendix}

\begin{figure*}
    \centering
    \includegraphics[width=0.96\linewidth]{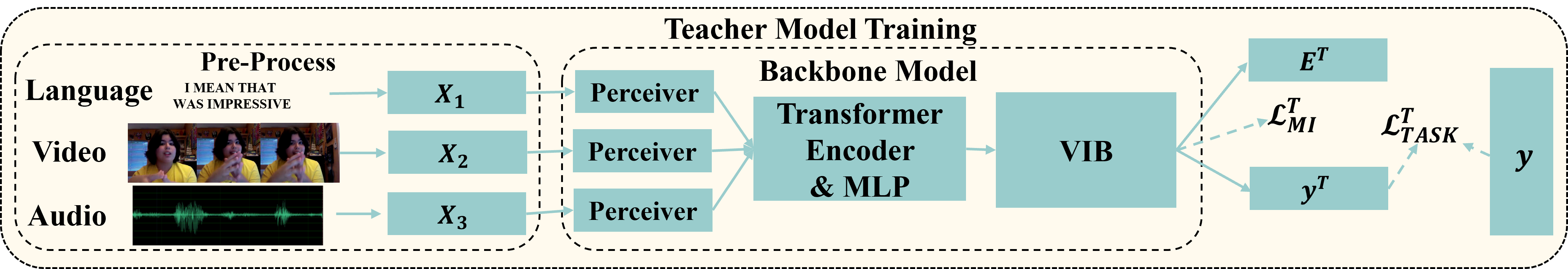}
    \caption{The structure of the teacher model.}
    \label{fig:teacher_model}
\end{figure*}

\subsection{Additional Model Details}
In this section, we provide the details of the proposed MCUR. The teacher and student models share the same architecture. As shown in Figure \ref{fig:teacher_model}, the teacher model consists of two main components: pre-processing and the backbone model, which we will explain in the following subsections.
\subsubsection{Pre-processing}
Given a multimodal dataset with $N$ samples $\mathcal{D}=\{X_1,X_2,...,X_m\}$, where $X_p\in\mathbb{R}^{t_p\times d_p}$ denotes the modality representations and $p\in\{1,2,...,m\}$ denotes the modality. Here $d_p$ and $t_p$ represent the dimensionality and sequence length for each modality, respectively. To standardize the dimensionality of the representations, we apply a 1D convolutional network with kernel size $3$ to process the input data, as follows:
\begin{equation}
X_p^w=W_{3}^w(X_p)
\end{equation}
where $X_p^w\in\mathbb{R}^{T\times D}$, $D$ and $T$ represent the common dimensionality and sequence length of all modalities, respectively. $W_{3\times 3}^w$ is the trainable weights, $w\in\{t,s\}$ indicates whether the representation is from the teacher ($t$) or student model ($s$).

\subsubsection{Backbone Model}\label{sec:backbone}
Once we obtain the modality representations $X^w_p$, the backbone model processes these representations to produce the final fused features and prediction logits. The backbone uses a Perceiver model \cite{jaegle2021perceiver, zhang2023learning, zhang2024towards}, which learns modality-specific representations by attending to input features using cross-modal transformer blocks.

In the cross-modal transformer block, the modality representations $X^w_p$ serve as keys and values, while learnable prompts $\hat{P^w_p} \in\mathbb{R}^{l_p\times D}$ act as queries. The attention mechanism is defined as:
\begin{equation}
P^w_p = \text{Softmax}(\frac{\hat{P^w_p}W_{Q_e}^w(W_{h}^w)^\top (X^w_p)^{\top}}{\sqrt{d_k}})X^w_pW_{V_h}^w,
\end{equation}
where $W_{Q_e}^w, (W_{h}^w)^\top, W_{V_h}^w \in\mathbb{R}^{D\times D}$ are trainable weights, and $l_p$ is the length of the prompt. After applying the cross-modal transformer, the resulting enhanced modality representation $P^w_p$ is obtained by averaging the outputs from the final transformer layer.

Once we have the processed representations $P^w_p$ from all $m$ modalities, the representations $P^w_p$ are then passed through a transformer encoder for interaction and then through a two-layer MLP for fusion, as follows:
\begin{equation}
E^w_f=W_2^w(W_1^w\text{TE}^w([P^w_1,P^w_2,..., P^w_m])+b_1^w)+b_2^w,
\end{equation}
where $\text{TE}^w(\cdot)$ is the two-layer transformer encoder, $E^w_f$ is the fused representation. $W_1^w$ and $W_2^w$ are trainable weights, $b_1^w$ and $b_2^w$ are biases. 

Since the $m$ modalities often contain redundant information, their combination can introduce noise that interferes with the effective transmission of information. To address this, the fused representation $E^w_f$ is passed through a Variational Information Bottleneck (VIB) \cite{gao2024embracing, alemi2022deep, tishby2000information, xiao2024neuro, mai2022multimodal} to reduce both redundancy and noise in the representations. VIB approximates the information bottleneck by learning an encoding $E^w$ that is maximally informative about the target $y$, while being compressive in relation to the fused input $E^w_f$. This is achieved by optimizing the following objective function:
\begin{equation}
\mathcal{L}_{VIB}^w=I(E^w,y)-\beta I(E^w,E^w_f),
\end{equation}
where $E^w$ is the compressed representation of $E^w_f$ and $y$ is the ground truth. $I(\cdot)$ represents mutual information, and $\beta$ is a Lagrange multiplier. In line with \cite{gao2024embracing}, the task loss $\mathcal{L}_{TASK}$ is integrated with Kullback-Leibler divergence to measure the mutual information terms $I(E^w,y)$ and $I(E^w,E^w_f)$ as expressed by:
\begin{equation}
\mathcal{L}_{VIB}^{w}=\mathcal{L}_{TASK}^w(y^{w},y)+\beta KL(p(e^w|e^w_f)||\mathcal{N}(0,\textbf{I})),
\end{equation}
where $p(e^w|e^w_f)\sim \mathcal{N}(\mu^w,(\sigma^w)^2\textbf{I})$, and the mean $\mu^w$ and standard deviation $\sigma^w$ are encoded using MLP layers through: $\mu^w=W_3^we^w_f+b_3^w$ and $\sigma^w=W_4^we^w_f+b_4^w$. $\mathcal{L}_{TASK}^w$ is introduced in Equations \ref{TASK1} and \ref{TASK2}. Using the re-parameterization \cite{kingma2013auto} trick $e^w=\mu^w+\epsilon \sigma^w$, where $\epsilon\sim \mathcal{N}(0,\textbf{I})$, the final representation $e^w$ is used to produce the logits $y^w$ through MLP layer: 
\begin{equation}
y^w=W_5^we^w+b_5^w.\label{y_c}
\end{equation}
Here $W_3^w$, $W_4^w$ and $W_5^w$ are trainable weights, $b_3^w$, $b_4^w$ and $b_5^w$ are biases in MLP. Note that all $w\in\{t,s\}$ indicates whether the elements are from the teacher ($t$) or student model ($s$).

\subsection{Additional Implementation Details}\label{sec:addi_imple}
This section outlines the training setup and implementation details, including the experiment setups, teacher model training details, modality setup and training strategy, baseline reproduction details.

\subsubsection{Missing Protocol Setups}
In this work, we evaluate and compare different models under two missing modality protocols.

The fixed missing protocol is designed to simulate realistic situations where a specific modality is consistently unavailable \cite{zhuang2025cmad,zhuang2025hyper,wei2023mmanet,li2024correlation,guo2024multimodal, wang2023distribution}. For instance, an audio sensor might fail, leaving only textual and visual information available. Evaluating models under this setting allows us to analyze their robustness to modality-specific failures that may occur during deployment.

In contrast, the random missing protocol reflects more common real-world conditions where missing data occurs unpredictably \cite{zhuang2025hyper,wang2023distribution,wang2023incomplete, zhang2024towards}. Factors such as poor signal quality or transmission errors can cause different modalities to be randomly missing across samples. We simulate this behavior by randomly dropping modalities in each sample to assess model performance under stochastic modality loss.

\subsubsection{Teacher Model Training Details}
The teacher model is trained using the following loss function:
\begin{equation}
\mathcal{L}_{all}=\mathcal{L}^{t}_{TASK}+\beta\mathcal{L}_{VIB}^{t}.
\end{equation}
For student model, the total loss is defined as:
\begin{equation}
\mathcal{L}_{all}=\gamma\mathcal{L}_{CL}^{s}+\beta\mathcal{L}_{VIB}^{s}+\mathcal{L}^{MSE}+\zeta \mathcal{L}^{Sugr},
\end{equation}
where $\beta$, $\gamma$ and $\zeta$ are hyper-parameters.

\subsubsection{Training Configuration} We conducted all experiments using PyTorch with CUDA 11.5 on a single NVIDIA RTX 3090 GPU. For optimization, we used AdamW \cite{loshchilov2018decoupled} on the MOSI and MOSEI datasets, and Adam \cite{kingma2014adam} for IEMOCAP. The teacher model was trained using complete modality inputs with a fixed $\beta$ value of 0.01. We set the learning rate to 2e-5 and used a random seed of 5576 for both MOSI and MOSEI. For IEMOCAP, the learning rate was set to 1e-3 with a random seed of 1111. The student model was trained using the same random seeds as the teacher model. The $\mu_1$ and $\mu_2$ are set as 1.0 by default. During student model's training, we performed hyper-parameter tuning with the learning rate searched over \{1e-5, 2e-5, 4e-5, 8e-4, 1e-3\}. The trade-off hyper-parameters $\gamma$ and $\zeta$ were searched over \{0.1, 0.2, 0.5, 1.0, 2.0, 5.0\} and \{1.0, 2.0, 5.0, 10.0, 20.0, 100.0\}, respectively. Finally, we trained MCUR on the MOSI dataset using a learning rate of 2e-5 with AdamW \cite{loshchilov2018decoupled}, $\gamma=2.0$, $\beta=0.01$, and $\zeta=1.0$. For the MOSEI dataset, we used a learning rate of 2e-5, $\gamma=0.2$, $\beta=0.01$, and $\zeta=5.0$, while for the IEMOCAP dataset, we set the learning rate to 8e-4 with Adam \cite{kingma2014adam}, $\alpha=0.2$, $\beta=0.01$, $\gamma=0.1$, $\zeta=100$ and $\tau=0.2$. 

\subsubsection{Modality Setup and Training Strategy} All KD-based methods, including MCUR, share the same teacher model, which is pretrained using complete modality inputs. For reconstruction-based approaches, all modalities are also available during training to guide the recovery of missing ones. To simulate modality loss during training, we uniformly sample from seven predefined missing modality patterns, each with a probability of 1/7.

\subsubsection{Baseline Reproduction Details}
Most existing methods are trained under the setting in which separate models are learned for different missing modality configurations \cite{wang2023distribution,wang2023incomplete,zhang2024towards}. While effective under controlled assumptions, this strategy often leads to limited robustness when modality availability varies in practice. To ensure a fair comparison and to better assess performance under conditions closer to real world scenarios, we reproduced all baseline methods using their official implementations whenever available.

For KD-based approaches such as CorrKD and MMANet, we adopted the same teacher model used in our MCUR framework to ensure consistency across methods. For approaches that require finetuning from pretrained checkpoints, including IMDer, we initialized the model using the official released checkpoints. All hyperparameters were set according to the configurations reported in the original papers and their corresponding open source code. When certain implementation details were not explicitly specified, such as the teacher model configuration for knowledge distillation methods, we aligned these settings with those used in MCUR to maintain fairness and reproducibility. Since the official code for CorrKD is not publicly available, we reimplemented the method based on the descriptions provided in the original paper. We first verified that our implementation achieved results comparable to those reported under the original experimental settings, and then applied it to the missing modality scenarios considered in this work. In cases where reproduced results differed from reported numbers, we conducted careful re-tuning to ensure competitive performance.

In the reproducibility experiments presented in Table \ref{tbl_generation} in Section \ref{sec_4_5_4}, we only added or modified the code corresponding to the evaluated components. The hyper-parameter search space was kept consistent with that used in MCUR, ensuring a fair and reliable comparison: $\gamma \in \{0.1, 0.2, 0.5, 1.0, 2.0, 5.0\}$, $\zeta \in \{1.0, 2.0, 5.0, 10.0, 20.0, 100.0\}$.

\subsubsection{Uncertainty Implementation Details}
In this section, we introduce the way to measure the uncertainty in Table \ref{abla_datasets} and Figure \ref{fig_UNCER}. For classification tasks, uncertainty is computed as entropy over predicted class probabilities:
\begin{equation}
H(p) = - \sum_{c} p_c \log p_c,
\end{equation}
where $p_c$ is obtained via softmax function.

Although MOSEI is trained as a regression task, we convert outputs into probabilities for uncertainty evaluation: $p = \sigma(z)$ where where $z$ is the model output and $\sigma$ is the sigmoid function. Then the brier score is calculated as: 
\begin{equation}
\text{Brier} = \frac{1}{N} \sum_{i=1}^{N} (p_i - y_i)^2.
\end{equation}
This corresponds exactly to: (1) applying sigmoid to logits and (2) computing mean squared error with binary labels. For NLL score, we use the following formula:
\begin{equation}
\text{NLL} = - \frac{1}{N} \sum_{i=1}^{N}\left[ y_i \log \sigma(z_i) + (1 - y_i)\log(1 - \sigma(z_i)) \right].
\end{equation}

\subsection{Additional Details of Datasets and Evaluation Metrics}
\textbf{Datasets}. In line with previous work \cite{guo2024multimodal, li2024correlation}, we evaluate the MCUR framework on three widely used MER datasets: MOSI \cite{zadeh2016mosi}, MOSEI \cite{zadeh2018multimodal}, and IEMOCAP \cite{busso2008iemocap}. Both MOSI and MOSEI are regression tasks, consisting of user-generated monologue videos collected from social media platforms such as YouTube. MOSI contains 2,199 video clips, with 1,284 for training, 229 for validation, and 686 for testing. MOSEI includes 22,856 video clips, with 16,326 for training, 1,871 for validation, and 4,659 for testing. Each video clip is annotated with a emotion polarity score ranging from -3 (strong negative) to 3 (strong positive). IEMOCAP \cite{busso2008iemocap}, on the other hand, is a multimodal dialogue dataset for classification task. It contains 4,453 video clips, with each clip labeled as one of the following four emotional classes: Neutral, Happy, Sad, and Angry.

\textbf{Evaluation Metrics}.
Following common practice \cite{guo2024multimodal, lian2023gcnet, li2024correlation,zhuang2025cmad}, for the MOSI and MOSEI datasets, we use accuracy (ACC) and F1 score computed for `positive/negative' classification results as evaluation metrics. For IEMOCAP, we calculate the accuracy and F1 score for each of the four categories and then compute their weighted average for evaluation.

\textbf{Feature Extraction}. For the language modality, we use pre-trained BERT embeddings \cite{devlin2019bert} for the MOSI and MOSEI datasets, and GLoVe word embeddings \cite{pennington2014glove} for the IEMOCAP dataset. For the video modality, we extracted facial features using the Facet toolkit \cite{baltruvsaitis2016openface} across all datasets. Audio features are extracted using the COVAREP toolkit \cite{degottex2014covarep} for all datasets.

\subsection{Baselines}
We compare MCUR with several state-of-the-art models, including both modality reconstruction-based and KD-based models, to provide a thorough evaluation.

\subsubsection{KD-based Methods}
\textbf{CorrKD} \cite{li2024correlation}: CorrKD employs a sample-level contrastive distillation mechanism for knowledge transfer across samples, addressing missing semantics. It aligns feature distributions using category prototypes and optimizes emotion decision boundaries via response disentanglement and mutual information maximization. 

\textbf{MMANet} \cite{wei2023mmanet}: MMANet offers a three-component framework: a deployment network, a teacher network, and a regularization network. It introduces margin-aware distillation module, which adjusts sample contributions based on classification uncertainty, and modality-aware regularization module to refine weak modality representations.

\subsubsection{Modality Reconstruction-based Methods}
\textbf{MPLMM} \cite{guo2024multimodal}: MPLMM utilizes three prompt types, including generative, missing-signal, and missing-type, to reconstruct missing modality features and enhance intra- and inter-modality learning. These prompts are used to reconstruct the missing modalities with available modalities.

\textbf{IMDer} \cite{wang2023incomplete}: IMDer recovers missing data by mapping random noise to the distribution space of the missing modality. It leverages available modalities as prior conditions to refine the recovery process, ensuring semantic alignment. 

\textbf{LNLN} \cite{zhang2024towards}: LNLN focuses on the language modality as the dominant emotion source. Its Dominant Modality Correction module and Dominant Modality-based Multimodal Learning modules enhance robustness against noise by improving dominant modality representations.

\subsection{Additional Results}\label{sec:additional_results}
In this section, we present additional experiments to further validate the effectiveness of MCUR.

First, we report the performance of MCUR under different missing-modality conditions on the MOSI dataset (Table \ref{EXPERI_MOSI}). We then provide the mean and variance across multiple runs on all three datasets to assess result stability (Table \ref{tbl_multi_run}). To further examine consistency, we report the mean and variance for each missing-modality setting over five random runs on the MOSEI and IEMOCAP datasets (Table \ref{EXPERI_STATIS}). Next, we evaluate the robustness of MCUR in more realistic settings (Table \ref{tbl_realworld}) and analyze its time and space complexity across datasets (Table \ref{UNCER_complexity}). We also report the mean and variance of prediction uncertainty for each missing-modality condition on all three datasets (Table \ref{EXPERI_UNCER}).

In addition, we conduct a detailed hyper-parameter analysis to understand the sensitivity of MCUR (Tables \ref{hyper-para} and \ref{hyper-para_mu}). We further evaluate the model under more challenging conditions where noise and missing modalities occur simultaneously (Table \ref{noise_results}). To better understand the behavior of MCUR, we analyze whether the student model inherits biases from the teacher model (Tables \ref{tab:teacher_effect} and \ref{tab:calibration_av}), and examine whether the learned representations are unified or dispersed. We also study the cross-sample generalization ability of MCUR (Table \ref{cross-dataset}).

Finally, we visualize the training loss to illustrate the optimization process (Figure \ref{fig_losses}), and present visualizations of the learned representations (Figure \ref{fig_URRL_VISUAL}).

\begin{table*}[ht]
\centering
\caption{Performance comparison in MOSI. `ACC/F1' is reported. Float values in `Avail' are MR.}\label{EXPERI_MOSI}
\begin{tabular}{ccccccc}
\hline
Avail & CorrKD & MPLMM & IMDer & LNLN & MMANet & MCUR \\
\hline
\hline
\multicolumn{7}{c}{Results on MOSI} \\
\hline
L & 83.38/83.22 & 82.77/82.59 & 83.23/83.30 & \textbf{84.60/84.56} & \textbf{84.60}/84.52 & 83.99/83.97 \\
V & 53.81/53.10 & 42.23/25.07 & 42.23/25.07 & 42.23/25.07 & 55.79/54.02 & \textbf{59.15/59.13} \\
A & 43.45/27.93 & 42.84/26.64 & 42.23/25.07 & 42.23/25.07 & 55.18/54.02 & \textbf{60.82/58.76} \\
L,V & 83.54/83.35 & 83.23/83.03 & 82.93/83.00 & 84.60/84.56 & 82.32/82.02 & \textbf{84.76/84.74} \\
L,A & 83.38/83.20 & 82.77/82.60 & 84.15/84.17 & 84.60/84.56 & 82.16/81.80 & \textbf{85.21/85.13} \\
A,V & 56.86/56.93 & 42.99/26.71 & 42.23/25.07 & 42.23/25.07 & \textbf{59.76/59.86} & 58.54/56.78 \\
L,A,V & 83.54/83.34 & 83.23/83.04 & 83.38/83.42 & 84.60/84.56 & 81.86/81.28 & \textbf{85.37/85.28} \\
0.1 & 81.25/81.19 & 80.95/80.93 & 81.25/81.36 & 80.64/80.73 & 80.49/80.02 & \textbf{82.62/82.50} \\
0.2 & 77.90/77.89 & 74.85/75.00 & 75.76/75.88 & 74.85/74.98 & 78.96/78.76 & \textbf{80.03/79.81} \\
0.3 & 73.63/73.76 & 71.80/71.88 & 75.30/75.36 & 71.04/70.91 & 74.54/74.55 & \textbf{78.96/78.65} \\
0.4 & 71.04/71.21 & 65.09/64.64 & 67.23/66.69 & 69.51/69.18 & 71.19/71.22 & \textbf{75.30/74.96} \\
0.5 & 67.84/67.94 & 62.96/62.05 & 57.32/54.04 & 61.43/59.61 & \textbf{70.43/70.58} & 70.27/69.76 \\
0.6 & 61.59/61.17 & 58.38/55.68 & 58.69/55.16 & 58.69/55.94 & 66.62/66.79 & \textbf{69.36/69.03} \\
0.7 & 60.21/59.24 & 53.20/49.01 & 59.30/55.98 & 57.47/54.17 & 63.87/63.98 & \textbf{67.99/67.42} \\
Avg. & 70.10/68.82 & 66.24/62.06 & 66.80/62.40 & 67.05/62.78 & 71.98/71.79 & \textbf{74.46/73.99} \\
\hline
\end{tabular}
\end{table*}

\subsubsection{Results on MOSI Dataset}
In this section, we analyze the performance of MCUR across three benchmark datasets. As shown in Table \ref{EXPERI_MOSI}, MCUR achieves consistently better results on the MOSI dataset under most modality-missing conditions, improving the average ACC and F1 by 2.48 and 2.20, respectively.

Although randomness introduces stochasticity that causes some performance variation across datasets (Table \ref{tbl_multi_run}), MCUR still delivers strong results overall. To better understand the effect of randomness, we report the mean and variance of prediction results under 14 different modality-missing scenarios on the MOSEI and IEMOCAP datasets (Table \ref{EXPERI_STATIS}). We observe that the fluctuations in F1 are generally higher than those in accuracy, but the average variation across all missing-modality settings is much smaller than that in any single condition. This finding suggests that MCUR can effectively handle diverse missing-modality situations and maintain consistent, robust performance.

\begin{table}{}
\centering
\caption{Statistics of MCUR with 5 different random seeds (95\% confidence intervals). Averaged `ACC/F1' of all missing scenarios is reported.}\label{tbl_multi_run}
\begin{tabular}{cccc}
\hline
& MOSI & MOSEI & IEMOCAP \\
\hline
ACC & 73.50$\pm$0.57 & 77.05$\pm$0.18 & 79.56$\pm$0.10 \\
F1 & 73.20$\pm$0.50 & 76.48$\pm$0.21 & 77.65$\pm$0.11 \\
\hline
\end{tabular}
\end{table}

\begin{table*}[ht]
\centering
\caption{Statistics of MCUR on datasets with 95\% confidence intervals.}\label{EXPERI_STATIS}
\begin{tabular}{cccccccc}
\hline
\multicolumn{8}{c}{Results on MOSEI} \\
\hline
 & L & V & A & L,V & L,A & A,V & L,A,V \\
\hline
ACC & 85.8$\pm$0.3 & 64.7$\pm$0.6 & 63.3$\pm$0.2 & 86.1$\pm$0.2 & 85.8$\pm$0.3 & 64.9$\pm$0.5 & 86.0$\pm$0.1 \\
F1 & 85.7$\pm$0.3 & 64.0$\pm$0.7 & 60.2$\pm$1.1 & 85.9$\pm$0.2 & 85.7$\pm$0.3 & 64.3$\pm$0.4 & 85.9$\pm$0.1 \\
\hline
MR & 0.1 & 0.2 & 0.3 & 0.4 & 0.5 & 0.6 & 0.7 \\
\hline 
ACC & 83.7$\pm$0.2 & 81.8$\pm$0.4 & 79.6$\pm$0.3 & 77.6$\pm$0.3 & 75.4$\pm$0.9 & 72.8$\pm$0.6 & 71.5$\pm$1.1 \\
F1 & 83.5$\pm$0.2 & 81.5$\pm$0.4 & 79.3$\pm$0.4 & 77.2$\pm$0.3 & 74.9$\pm$1.0 & 72.1$\pm$0.6 & 70.6$\pm$0.9 \\
\hline
\hline
\multicolumn{8}{c}{Results on IEMOCAP} \\
\hline
& L & V & A & L,V & L,A & A,V & L,A,V \\
\hline
ACC & 78.9$\pm$0.5 & 75.4$\pm$0.1 & 79.8$\pm$0.3 & 79.3$\pm$0.4 & 81.6$\pm$0.5 & 80.0$\pm$0.7 & 81.8$\pm$0.1 \\
F1 & 77.3$\pm$0.5 & 70.5$\pm$0.5 & 77.3$\pm$0.5 & 77.8$\pm$0.4 & 80.4$\pm$0.5 & 78.1$\pm$0.7 & 80.8$\pm$0.2 \\
\hline
MR & 0.1 & 0.2 & 0.3 & 0.4 & 0.5 & 0.6 & 0.7 \\
\hline
ACC & 81.6$\pm$0.4 & 81.1$\pm$0.7 & 80.4$\pm$0.5 & 79.9$\pm$0.5 & 78.6$\pm$0.4 & 78.3$\pm$0.5 & 77.7$\pm$0.6 \\
F1 & 80.5$\pm$0.5 & 79.4$\pm$0.3 & 79.0$\pm$0.6 & 78.3$\pm$0.6 & 76.6$\pm$0.5 & 76.0$\pm$0.6 & 75.2$\pm$0.7 \\
\hline
\end{tabular}
\end{table*}

\subsubsection{Robustness in Real-World Scenarios}
To comprehensively assess the robustness of the MCUR framework in real-world scenarios, we conducted additional experiments in this section. Since modality loss in real-life situations is inherently uncertain, we simulated this uncertainty by running the trained model ten times on the test set with different random seeds. MCUR's average performance with 95\% confidence intervals across 14 modality missing conditions in three datasets is reported in Table \ref{tbl_realworld}.

Our findings indicate that MCUR effectively handles various modality missing situations encountered in real-world scenarios. The results across all three datasets showed low variance, highlighting MCUR's stability and robustness. Specifically, in the MOSEI dataset, the fluctuations in `ACC/F1' scores were less than 0.2, even under the most challenging conditions. Although the MOSI dataset demonstrated slightly higher fluctuations, the performance variation did not exceed 0.5. These results further demonstrate the superior robustness and reliability of MCUR.

\begin{table}{}
\centering
\caption{Robustness of MCUR in real-world Scenarios with 10 different random seeds (95\% confidence intervals). Averaged `ACC/F1' of all missing scenarios is reported.}\label{tbl_realworld}
\begin{tabular}{cccc}
\hline
& MOSI & MOSEI & IEMOCAP \\
\hline
ACC & 74.46$\pm$0.21 & 77.16$\pm$0.08 & 79.64$\pm$0.11 \\
F1 & 74.04$\pm$0.22 & 76.73$\pm$0.09 & 77.60$\pm$0.11 \\
\hline
\end{tabular}
\end{table}

\begin{table}[t]
\centering
\caption{Complexity of the MCUR.}\label{UNCER_complexity}
\begin{tabular}{ccccc}
\hline
Datasets & \#params & GPU Memory & Time \\
 \hline
MOSI & 112,627,841 & 5,542.6MB & 686.5s \\
MOSEI & 116,421,953 & 6,234.6MB & 7,660.1s \\
IEMOCAP & 624,654 & 1,166.6MB & 1,194.8s \\
\hline
\end{tabular}
\end{table}

\subsubsection{Computation Overhead}
To assess the computational efficiency of MCUR, we trained the model for 100 epochs on an NVIDIA GTX 3090 GPU and recorded its number of parameters, peak GPU memory usage, and training time across the three datasets (Table \ref{UNCER_complexity}).

We found that MCUR requires the largest number of parameters and computational resources on the MOSEI dataset. However, even in this case, the peak GPU memory usage was only 6.2 GB, allowing MCUR to run comfortably on consumer-grade GPUs. Training on MOSEI completed in approximately two hours, demonstrating the practicality and efficiency of the proposed framework.

\begin{table*}[ht]
\centering
\caption{Statistics of prediction uncertainty on datasets with 95\% confidence intervals.}\label{EXPERI_UNCER}
\begin{tabular}{cccccc}
\hline
\hline
\multicolumn{6}{c}{Results on MOSI} \\
\hline
 & L & V & A & L,V & L,A \\
\hline
Brier & 0.154$\pm$0.030 & 0.254$\pm$0.003 & 0.249$\pm$0.001 & 0.155$\pm$0.030 & 0.153$\pm$0.030 \\
NLL & 0.482$\pm$0.067 & 0.701$\pm$0.005 & 0.691$\pm$0.001 & 0.484$\pm$0.066 & 0.480$\pm$0.068 \\
\hline
MR & 0.1 & 0.2 & 0.3 & 0.4 & 0.5 \\
\hline 
Brier & 0.163$\pm$0.026 & 0.174$\pm$0.023 & 0.183$\pm$0.020 & 0.193$\pm$0.016 & 0.203$\pm$0.016 \\
NLL & 0.502$\pm$0.057 & 0.526$\pm$0.050 & 0.546$\pm$0.045 & 0.569$\pm$0.035 & 0.589$\pm$0.036 \\
\hline
MR & 0.6 & 0.7 & A,V & L,A,V &  \\
\hline 
Brier & 0.213$\pm$0.012 & 0.219$\pm$0.012 & 0.250$\pm$0.001 & 0.154$\pm$0.030 & \\
NLL & 0.611$\pm$0.027 & 0.625$\pm$0.026 & 0.693$\pm$0.002 & 0.481$\pm$0.067 & \\
\hline
\hline
\multicolumn{6}{c}{Results on MOSEI} \\
\hline
& L & V & A & L,V & L,A \\
\hline
Brier & 0.143$\pm$0.003 & 0.231$\pm$0.002 & 0.239$\pm$0.001 & 0.142$\pm$0.002 & 0.143$\pm$0.003 \\
NLL & 0.462$\pm$0.007 & 0.654$\pm$0.005 & 0.671$\pm$0.003 & 0.457$\pm$0.005 & 0.461$\pm$0.006 \\
\hline
MR & 0.1 & 0.2 & 0.3 & 0.4 & 0.5 \\
\hline
Brier & 0.150$\pm$0.002 & 0.159$\pm$0.002 & 0.168$\pm$0.002 & 0.178$\pm$0.002 & 0.187$\pm$0.003 \\
NLL & 0.477$\pm$0.004 & 0.496$\pm$0.005 & 0.516$\pm$0.004 & 0.537$\pm$0.005 & 0.557$\pm$0.006 \\
\hline
MR & 0.6 & 0.7 & A,V & L,A,V &  \\
\hline
Brier & 0.198$\pm$0.002 & 0.205$\pm$0.003 & 0.227$\pm$0.002 & 0.141$\pm$0.002 & \\
NLL & 0.582$\pm$0.004 & 0.597$\pm$0.006 & 0.647$\pm$0.005 & 0.457$\pm$0.004 & \\
\hline
\hline
\multicolumn{6}{c}{Results on IEMOCAP} \\
\hline
& L & V & A & L,V & L,A \\
\hline
Brier & 0.166$\pm$0.006 & 0.181$\pm$0.003 & 0.158$\pm$0.009 & 0.162$\pm$0.006 & 0.149$\pm$0.010 \\
NLL & 0.509$\pm$0.016 & 0.546$\pm$0.008 & 0.489$\pm$0.023 & 0.499$\pm$0.016 & 0.468$\pm$0.025 \\
\hline
MR & 0.1 & 0.2 & 0.3 & 0.4 & 0.5 \\
\hline
Brier & 0.149$\pm$0.009 & 0.151$\pm$0.009 & 0.153$\pm$0.009 & 0.159$\pm$0.008 & 0.160$\pm$0.008 \\
NLL & 0.467$\pm$0.022 & 0.473$\pm$0.023 & 0.478$\pm$0.023 & 0.493$\pm$0.019 & 0.495$\pm$0.020 \\
\hline
MR & 0.6 & 0.7 & A,V & L,A,V & \\
\hline
Brier & 0.166$\pm$0.006 & 0.169$\pm$0.005 & 0.155$\pm$0.009 & 0.146$\pm$0.010 & \\
NLL & 0.509$\pm$0.015 & 0.518$\pm$0.012 & 0.482$\pm$0.023 & 0.459$\pm$0.025 & \\
\hline
\end{tabular}
\end{table*}

\subsubsection{Additional Uncertainty Analysis}
In this section, we present the mean and variance of prediction uncertainty for each missing-modality condition across all three datasets in Table \ref{EXPERI_UNCER}, corresponding to the complete results of Figure \ref{fig_UNCER}.

Under the random missing protocol, uncertainty increases monotonically with the MR. On MOSEI, uncertainty rises quickly but remains stable across different runs, while on IEMOCAP, it grows more gradually yet exhibits higher variance. Under the fixed missing protocol, similar trends are observed, though the uncertainty distribution depends on which modality is missing. Consistent with the random setting, MOSEI shows smaller fluctuations, whereas IEMOCAP displays larger ones. The absolute uncertainty also varies more significantly across missing-modality types. On MOSEI, retaining the `A', `V', or `A,V' modalities leads to noticeably higher uncertainty, reflecting the dominant predictive role of the `L'. In contrast, on IEMOCAP, where different modalities contribute more evenly, the variation in uncertainty across missing-modality conditions is less pronounced.

\subsubsection{Hyper-parameter Analysis}
In this section, we conduct sensitivity analyses on four key hyper-parameters, namely $\gamma$, $\zeta$, $\mu_1$, and $\mu_2$, on MOSEI and IEMOCAP. The average performance over 14 missing modality settings is reported in Tables \ref{hyper-para} and \ref{hyper-para_mu}.

\begin{table}[htbp]
\centering
\caption{Performance of different $\gamma$ and $\zeta$ values. Averaged `ACC/F1' is reported.}\label{hyper-para}
\begin{tabular}{ccc|ccc}
\hline
$\gamma$ & MOSEI & IEMOCAP & $\zeta$ & MOSEI & IEMOCAP \\
\hline
0.1 & 76.6/76.1 & 79.7/77.7 & 1 & 76.7/76.2 & 78.8/76.5\\
0.2 & 77.2/76.8 & 79.3/77.2 & 2 & 77.0/76.3 & 78.8/76.6\\
0.5 & 76.9/76.1 & 79.0/76.5 & 5 & 77.2/76.8 & 78.8/76.5 \\
1 & 76.6/76.0 & 79.0/76.1 & 10 & 76.5/76.0 & 78.9/76.6 \\
2 & 75.9/75.8 & 78.9/75.3 & 20 & 76.3/76.0 & 79.2/77.1 \\
5 & 75.8/75.5 & 78.9/75.2 & 50 & 76.2/75.7 & 79.4/77.0 \\
10 & - & - & 100 & 75.8/75.7 & 79.7/77.7 \\
\hline
\end{tabular}
\end{table}

\textbf{Sensitivity of $\gamma$ and $\zeta$}. Table \ref{hyper-para} shows that MCUR performs best with relatively small values of $\gamma$. On MOSEI, the best result is obtained at $\gamma = 0.2$, with an ACC/F1 of 77.2/76.8, while on IEMOCAP the best result appears at $\gamma = 0.1$, with 79.7/77.7. As $\gamma$ increases, performance gradually declines on both datasets. This indicates that the alignment objective is helpful, but it should remain a supporting term rather than dominate the optimization. A moderate value of $\gamma$ therefore provides the best balance between representation alignment and emotion classification.

The effect of $\zeta$ is more stable on MOSEI but more pronounced on IEMOCAP. On MOSEI, the results stay within a narrow range, with the best performance at $\zeta = 5$. On IEMOCAP, larger values of $\zeta$ generally bring better results, and the best performance is achieved at $\zeta = 100$, with an ACC/F1 of 79.7/77.7. This suggests that stronger weighting of the SUGR module is especially helpful when the data are more sensitive to uncertainty caused by missing modalities. Overall, MCUR remains stable across a wide range of $\zeta$ values, while still benefiting from a larger value on the more challenging dataset.

Taken together, these results show that MCUR is not highly sensitive to the exact choice of $\gamma$ and $\zeta$. The performance changes smoothly rather than sharply, which indicates that both the contrastive objective and the uncertainty guided regularization provide reliable guidance under different settings.

\textbf{Sensitivity of $\mu_1$ and $\mu_2$}. As shown in Table \ref{hyper-para_mu}, performance consistently improves as $\mu_1$ increases from 0.2 to 1.0, where the best results are achieved on both datasets. Importantly, the neighboring settings $\mu_1 = 0.8$ and $\mu_1 = 1.2$ remain highly competitive, indicating that the model is robust to moderate variations of this parameter. Compared with $\mu_1 = 1.0$, a smaller value ($\mu_1 = 0.8$) slightly weakens semantic supervision and leads to a minor performance drop, but helps retain more modality-related information. In contrast, a larger value ($\mu_1 = 1.2$) strengthens category supervision and maintains comparable accuracy, while slightly degrading F1, suggesting a reduced benefit from modality-aware alignment. Overall, $\mu_1 = 1.0$ provides the best tradeoff, while nearby values confirm stable behavior around this optimum.

A similar trend is observed for $\mu_2$, where the best performance is obtained at $\mu_2 = 1.0$ on both datasets. The results at $\mu_2 = 0.8$ and $\mu_2 = 1.2$ remain close to the optimum, further demonstrating the robustness of MCUR. Specifically, a smaller value ($\mu_2 = 0.8$) relaxes the regularization and preserves more modality-specific details, at the cost of a slight performance decrease. A larger value ($\mu_2 = 1.2$) enforces stronger suppression of redundant information, which remains effective on IEMOCAP but leads to a mild degradation on MOSEI, indicating that excessive regularization may remove useful cues. These observations suggest that $\mu_2 = 1.0$ achieves a balanced tradeoff, while the model maintains stable performance within a reasonable range around this setting.

\begin{table}[htbp]
\centering
\caption{Performance of different $\mu_1$ and $\mu_2$ values. Averaged `ACC/F1' is reported.}\label{hyper-para_mu}
\begin{tabular}{ccc|ccc}
\hline
$\mu_1$ & MOSEI & IEMOCAP & $\mu_2$ & MOSEI & IEMOCAP \\
\hline
0.2 & 76.43/76.21 & 78.75/76.28 & 0.2 & 76.77/76.12 & 79.27/77.07 \\
0.4 & 76.90/76.28 & 78.85/76.33 & 0.4 & 76.89/76.19 & 79.34/77.49 \\
0.6 & 76.38/76.23 & 78.86/76.31 & 0.6 & 76.96/76.34 & 79.01/77.15 \\
0.8 & 77.01/76.43 & 79.46/77.63 & 0.8 & 77.06/76.42 & 79.64/77.65 \\
1.0 & 77.23/76.80 & 79.69/77.65 & 1.0 & 77.23/76.80 & 79.69/77.65 \\
1.2 & 77.03/76.42 & 79.70/77.58 & 1.2 & 76.77/76.38 & 79.44/77.54 \\
\hline
\end{tabular}
\end{table}

\subsubsection{Results on Noisy and Incomplete Settings}
Although MCUR does not explicitly distinguish between "missing" and "noisy" modalities, it measures sample reliability through teacher–student uncertainty discrepancy. Both missing and noisy inputs reduce effective information and increase uncertainty, and are thus handled in a unified way. To verify this, we extend our experiments to scenarios where both missing modalities and noisy inputs are present. 

Specifically, we inject Gaussian noise into each representation as in prior research \cite{zhuang2026tmdc} through:
\begin{equation}
embed = embed + intensity\cdot\mathcal{N}(0, \textbf{I}), 
\end{equation}
where $intensity$ controls the noise level.

The results are shown in Table \ref{noise_results}. As noise increases, all models degrade, but MCUR consistently remains the most robust. Under strong noise, e.g, $intensity=20$, MCUR achieves an F1 of 74.0, demonstrating clear noise resistance.

\begin{table}[htbp]
\centering
\caption{Performance of MCUR on noisy datasets. Averaged `ACC/F1' is reported.}\label{noise_results}
\begin{tabular}{c|cc|cc|cc}
\hline
Model & $intensity$ & ACC/F1 & $intensity$ & ACC/F1 & $intensity$ & ACC/F1 \\
\hline
MMANet & 5 & 75.0/74.3 & 10 & 74.7/73.4 & 20 & 74.4/72.2 \\
LNLN & 5 & 75.7/71.4 & 10 & 75.1/71.1 & 20 & 73.7/69.7 \\
MCUR & 5 & \textbf{75.9/74.6} & 10 & \textbf{75.4/74.4} & 20 & \textbf{74.6/74.0} \\
\hline
\end{tabular}
\end{table}

\subsubsection{Analysis of Fragmented Representation}\label{sec_fragment}
A potential concern is that conditioning the contrastive objective on modality combinations in MCB-CL may lead to fragmented representations, rather than a unified multimodal space. However, strict modality invariance, i.e., forcing all modality combinations to collapse into a single representation, is not well-suited to missing-modality settings. Different modality subsets (e.g., text-only vs. audio-visual) provide inherently heterogeneous and incomplete observations. Enforcing full alignment across such conditions can obscure modality-specific cues and introduce dominance effects from stronger modalities.

This limitation is also reflected in prior approaches that aim to learn unified representations. Despite this objective, the resulting embeddings are often biased toward the language modality. As a consequence, their performance degrades substantially when text is unavailable (Table \ref{EXPERI_all_fix}), suggesting that the learned representations are not truly modality-invariant in practice, but instead rely heavily on language signals.

In contrast, MCB-CL is designed to balance semantic consistency and modality awareness. It encourages representations to remain aligned at the semantic level across modality conditions, while preserving informative differences induced by varying modality subsets. This leads to a structured embedding space where global relationships are maintained without enforcing unnecessary collapse.

To further examine this property, we analyze the cosine similarity between teacher (full-modality) and student (missing-modality) representations. When text is present, the similarity is approximately 0.69. Under missing-text conditions (e.g., audio-visual only), similarity remains consistently high, ranging from 0.63 to 0.75, with all results statistically significant ($p<0.001$). In addition, we observe reduced intra-class variance, indicating that the representations become more compact while preserving their global organization. These observations suggest that MCB-CL does not fragment the embedding space. Instead, it promotes alignment with the full-modality semantic structure while improving representation compactness.

This behavior is also reflected in downstream performance. When text is available, the teacher model achieves slightly higher performance than the student (nearly 0.26\% F1), which is expected given its access to complete information. In contrast, under audio-visual-only conditions, the student outperforms the teacher by a substantial margin (15.8\% F1), indicating improved utilization of non-language modalities. 

Overall, both representation-level analysis and predictive performance consistently indicate that MCB-CL enhances robustness to missing modalities while maintaining coherent semantic structure, rather than inducing fragmented representations.

\subsubsection{Analysis of Transfer Bias from Teacher Model}
As discussed in Section~\ref{sec_fragment}, MCUR tends to rely more heavily on textual information in both the teacher and student models. We therefore examine how teacher quality and teacher bias affect student learning under missing-modality conditions.

To this end, we conduct controlled experiments in which only the teacher distribution is perturbed, while the student architecture, optimizer, training schedule, and all MCUR losses remain unchanged. Evaluation follows the same 14 missing-modality scenarios used in the main paper, including 7 fixed and 7 random settings.

To quantify language dominance, we define two auxiliary metrics. Let
\begin{equation}
S = \{L, V, A, L+V, L+A, A+V, L+A+V\}. 
\end{equation}
We define:
\begin{equation}
\text{LangDomGap} = F1(L) - \frac{1}{3}\big(F1(A) + F1(V) + F1(A+V)\big),
\end{equation}
which measures the performance advantage of language over non-language settings. Larger values indicate stronger language dominance.

We also define:
\begin{equation}
\text{BiasAmp} = \frac{\text{LangDomGap}_{\text{student}}}{\text{LangDomGap}_{\text{teacher}} + 1\mathrm{e}{-8}},
\end{equation}
where values below 1 indicate that the student attenuates the teacher's bias.

We consider four teacher settings with identical model architecture but different training perturbations: (1) \textit{noise}, which injects Gaussian noise into teacher embeddings ($tense=5$); (2) \textit{mr\_30}, which applies 30\% random modality dropout during teacher training; (3) \textit{high\_l}, which increases the sampling probability of the language-only setting to $P(L)=4/7$ and assigns $1/14$ to each of the remaining six combinations; and (4) \textit{high\_l\_2}, which biases sampling toward language-related combinations with probabilities $[1/7,1/14,1/14,3/14,3/14,1/14,3/14]$ over $[L,A,V,L+A,L+V,A+V,L+A+V]$. The results are shown in Table \ref{tab:teacher_effect}.

\begin{table}[t]
\centering
\caption{Effect of teacher quality and bias on student performance on MOSEI.}
\label{tab:teacher_effect}
\begin{tabular}{lccccccc}
\hline
Teacher & T Avg F1 & S Avg F1 & $\Delta$Avg & T Worst F1 & S Worst F1 & $\Delta$Worst & BiasAmp \\
\hline
$noise$      & 71.35 & 75.34 & \textbf{+3.99} & 48.51 & 57.14 & \textbf{+8.63} & 0.6350 \\
$mr\_30$          & 72.23 & 75.50 & \textbf{+3.26} & 48.51 & 59.38 & \textbf{+10.87} & 0.5810 \\
$high\_l$         & 72.28 & 75.66 & \textbf{+3.38} & 48.51 & 59.47 & \textbf{+10.96} & 0.6133 \\
$high\_l\_2$ & 72.48 & 75.31 & \textbf{+2.83} & 48.51 & 58.38 & \textbf{+9.87} & 0.6521 \\
\hline
\end{tabular}
\end{table}

Across all four teacher settings, MCUR consistently improves average F1 by 2.8 to 4.0 points and worst-case F1 by 8.6 to 11.0 points. The gains are largest in non-language settings. For example, under \textit{noise}, the student improves by 8.63 on $V$, 14.96 on $A$, and 15.86 on $A+V$; under \textit{mr\_30}, the corresponding gains are 10.86, 15.48, and 15.95. Similar patterns hold for \textit{high\_l} and \textit{high\_l\_2}, indicating that the performance gains mainly come from better use of non-language modalities rather than increased reliance on text.

This trend is also reflected in the bias statistics. LangDomGap decreases from 0.3558-0.3732 for the teachers to 0.2164-0.2434 for the students, corresponding to a reduction of 34.8\%-41.9\%. BiasAmp remains below 1 in all cases, showing that the student attenuates rather than amplifies the teacher's language bias. Similar reductions are also observed for ACC, with BiasAmp(ACC) consistently below 1.

\begin{table}[t]
\centering
\caption{Calibration comparison on non-language settings ($A$ and $A{+}V$). Lower is better.}
\label{tab:calibration_av}
\begin{tabular}{llcccc}
\hline
 &  & \multicolumn{2}{c}{A} & \multicolumn{2}{c}{A+V} \\
\cmidrule(lr){3-4} \cmidrule(lr){5-6}
Teacher & Model & Brier $\downarrow$ & NLL $\downarrow$ & Brier $\downarrow$ & NLL $\downarrow$ \\
\hline
\multirow{2}{*}{add\_noise} 
& Teacher  & 0.2702 & 0.7632 & 0.2712 & 0.7672 \\
& Student  & \textbf{0.2243} & \textbf{0.6398} & \textbf{0.2214} & \textbf{0.6337} \\
\hline
\multirow{2}{*}{mr\_30} 
& Teacher  & 0.2331 & 0.6590 & 0.2305 & 0.6532 \\
& Student  & \textbf{0.2286} & \textbf{0.6495} & \textbf{0.2265} & \textbf{0.6449} \\
\hline
\multirow{2}{*}{high\_l} 
& Teacher  & 0.2331 & 0.6590 & 0.2305 & 0.6533 \\
& Student  & \textbf{0.2289} & \textbf{0.6500} & \textbf{0.2286} & \textbf{0.6493} \\
\hline
\multirow{2}{*}{high\_l\_2} 
& Teacher  & 0.2331 & 0.6589 & 0.2305 & 0.6532 \\
& Student  & \textbf{0.2276} & \textbf{0.6473} & \textbf{0.2255} & \textbf{0.6429} \\
\hline
\end{tabular}
\end{table}

We further analyze calibration on the non-language settings ($A$ and $A+V$) in Table~\ref{tab:calibration_av}. Under the \textit{add\_noise} setting, MCUR yields substantial improvements in both Brier score and NLL. For example, on $A$ and $A+V$, Brier decreases from 0.2702/0.2712 to 0.2243/0.2214, and NLL decreases from 0.7632/0.7672 to 0.6398/0.6337, respectively. Under the remaining teacher settings (\textit{mr\_30}, \textit{high\_l}, and \textit{high\_l\_2}), MCUR consistently reduces both Brier score and NLL across $A$ and $A+V$. Although the magnitude of improvement is smaller compared to the noisy setting, the trend is stable across all perturbations.

These results indicate that MCUR not only improves predictive performance but also produces more reliable probability estimates in non-language scenarios. This is particularly important in missing-modality settings, where uncertainty is inherently higher. Together with the performance and bias analyses, the calibration results further support that the observed gains arise from improved robustness rather than over-reliance on language signals.

\begin{figure}[ht]
  \centering
  \includegraphics[width=0.47\linewidth]{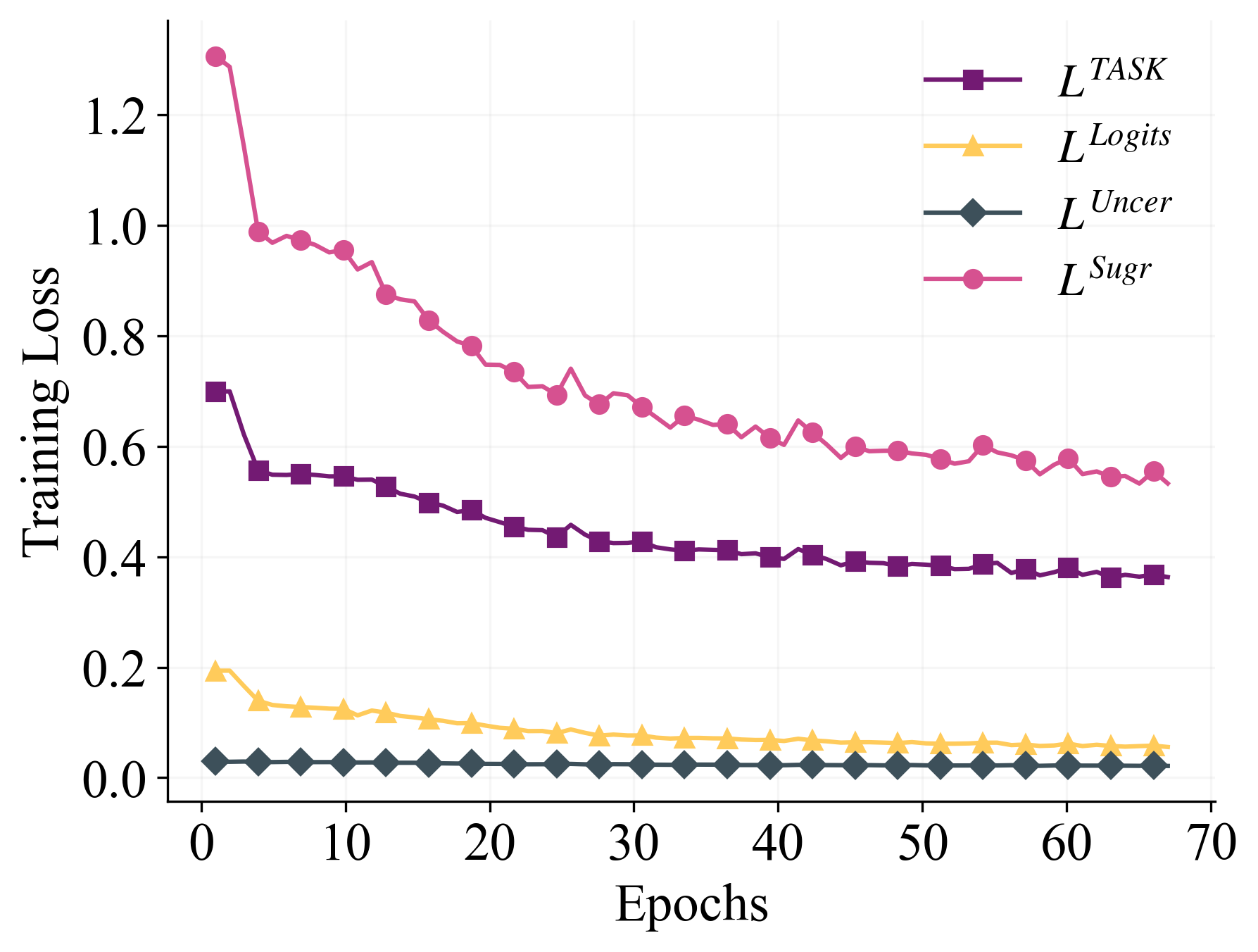}\label{fig_loss_ubr}
  \includegraphics[width=0.47\linewidth]{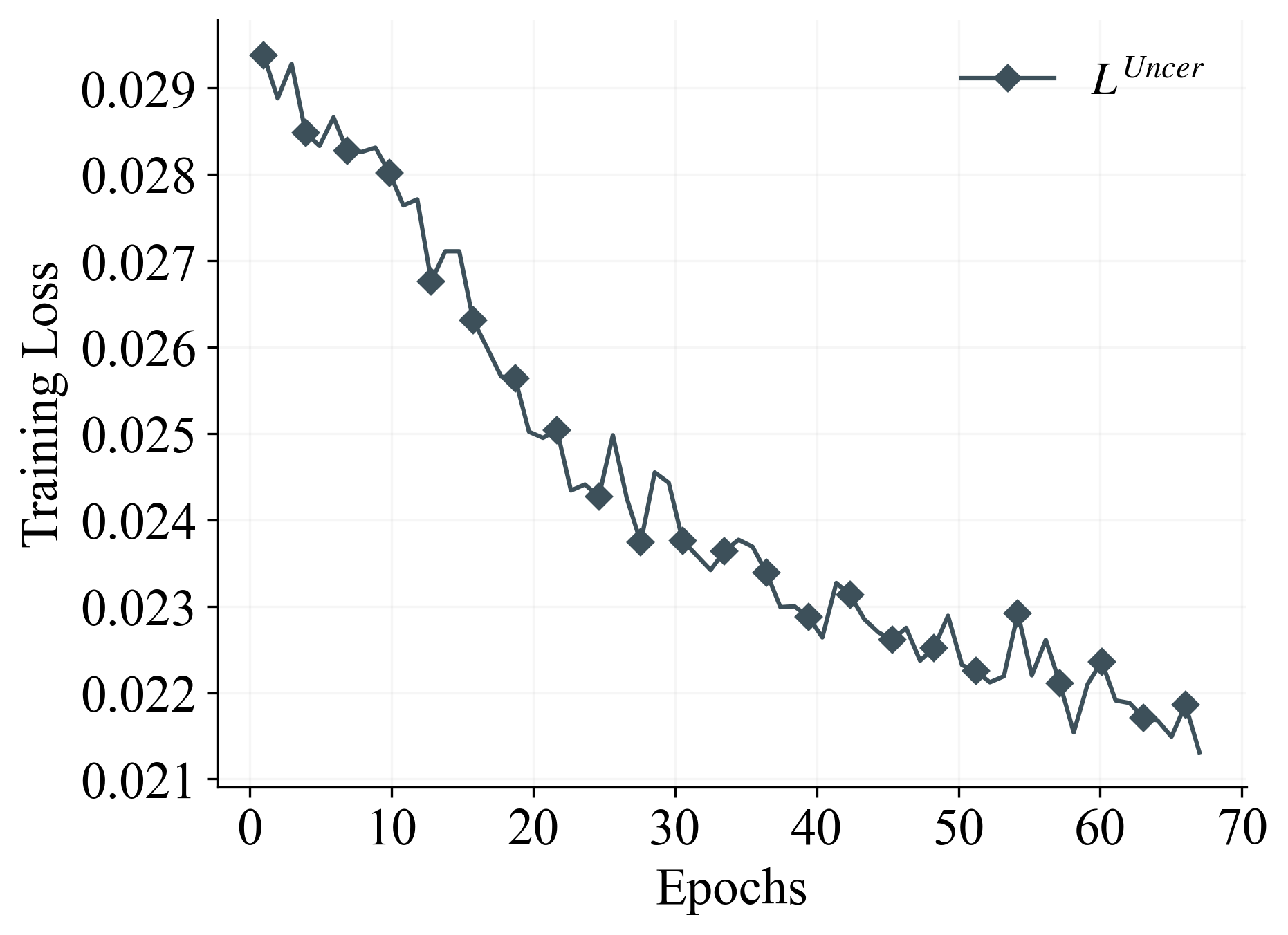}\label{fig_loss_uncertain}
 \caption{Analysis on the training convergence on IEMOCAP. (a) $L^{Sugr}$ Components; (b) $L^{Uncer}$.}\label{fig_losses}
\end{figure}

\subsubsection{Cross-Dataset Generalization}
In this section, we test the generalization ability of MCUR. Specifically, we train MCUR on MOSEI dataset and test on MOSI dataset for that these two datasets are all regression tasks. We use a strict transfer setting with no fine-tuning and only simple zero-padding to match the input shape. The results are shown in Table \ref{cross-dataset}.

\begin{table}[htbp]
\centering
\caption{Cross-dataset generalization results. Averaged `ACC/F1' is reported.}\label{cross-dataset}
\begin{tabular}{cc}
\hline
Model & Student \\
\hline
Shared Teacher Model & \textbf{76.14}/69.95 \\
CorrKD & 62.76/48.74 \\
MMANet & 74.88/71.68 \\
MCUR & 75.46/\textbf{74.55} \\
\hline
\end{tabular}
\end{table}

MCUR preserves teacher-level accuracy while improving F1 by 4.59 over the teacher and 2.87 over prior KD students, indicating that the learned consistency generalizes beyond a single dataset.

\subsubsection{Loss Analysis}
To better understand the internal behavior of the SUGR module, we analyze how its loss components evolve during training. Specifically, we visualize the training dynamics of each loss on the IEMOCAP dataset, as shown in Figure \ref{fig_losses}. Here, all loss components represent the average values across all representations at each epoch.

Our observations are as follows:
(1) The components of $L^{Sugr}$ consistently decrease over epochs, indicating that each part of the SUGR module is well-optimized and contributes effectively to training.
(2) In Figure \ref{fig_losses}(b), we examine $L^{Uncer}$, which captures uncertainty differences. While it shows minor fluctuations, the overall trend is downward. These fluctuations may result from inconsistencies in the missing modalities across epochs, causing stronger modalities to be lost at different rates in different batches. These results validate the design of the SUGR module and demonstrate its stable behavior during training under missing modality conditions.

\subsection{Visualization of the Representations}
\begin{figure*}[t]
\includegraphics[width=0.32\linewidth]{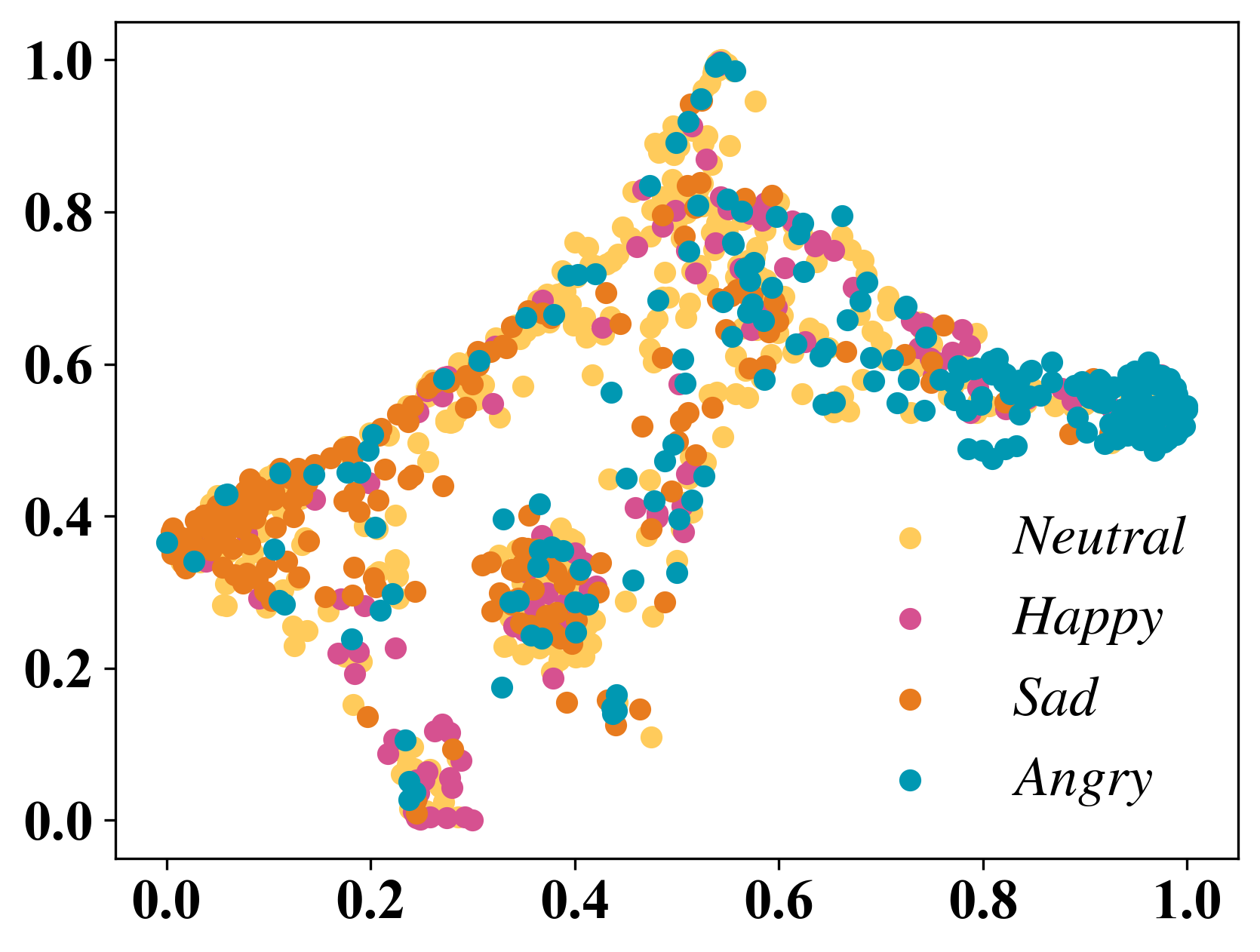}\label{fig_URRL_ALL} \hfill
\includegraphics[width=0.32\linewidth]{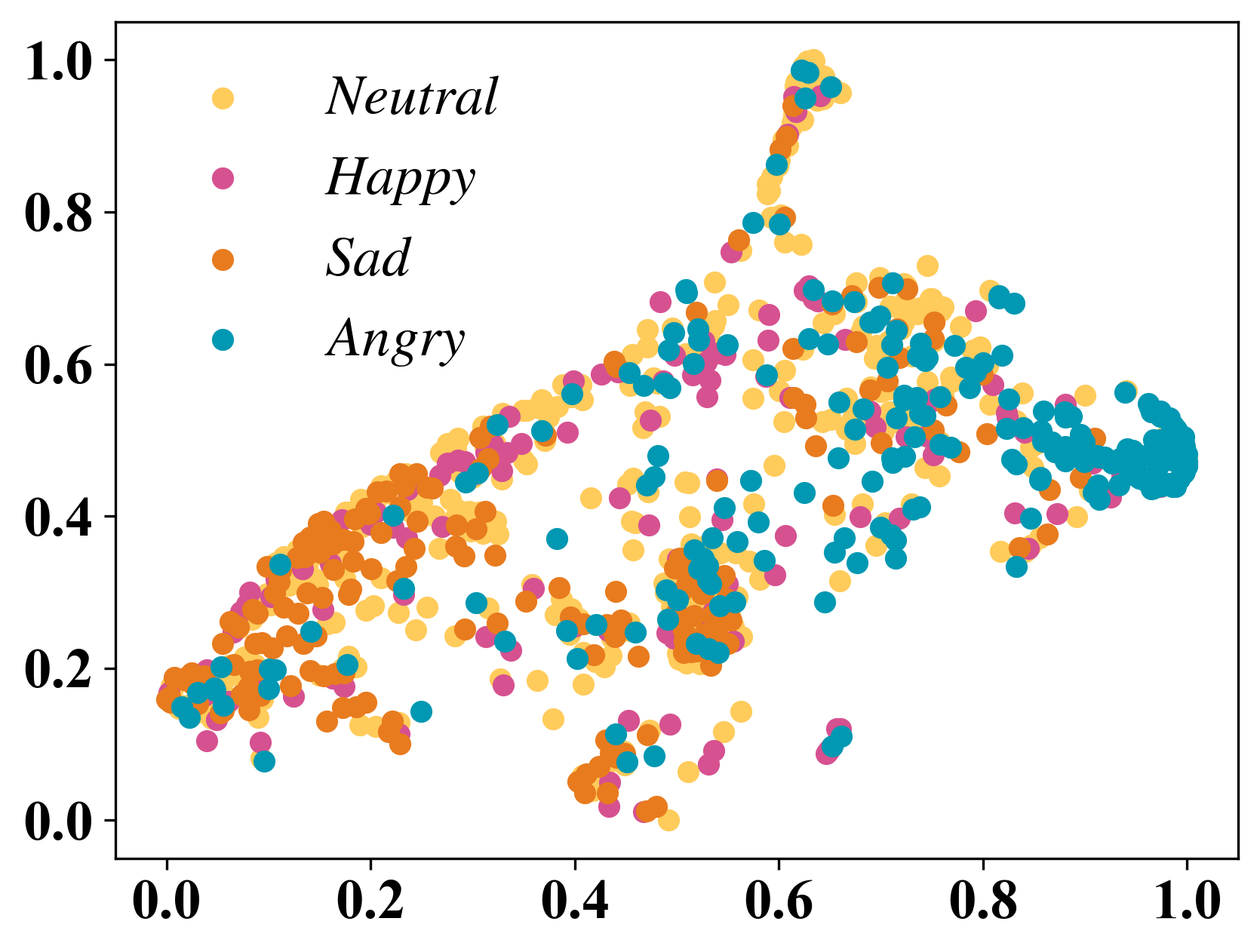}\label{fig_URRL_WO_CL}\hfill
\includegraphics[width=0.32\linewidth]{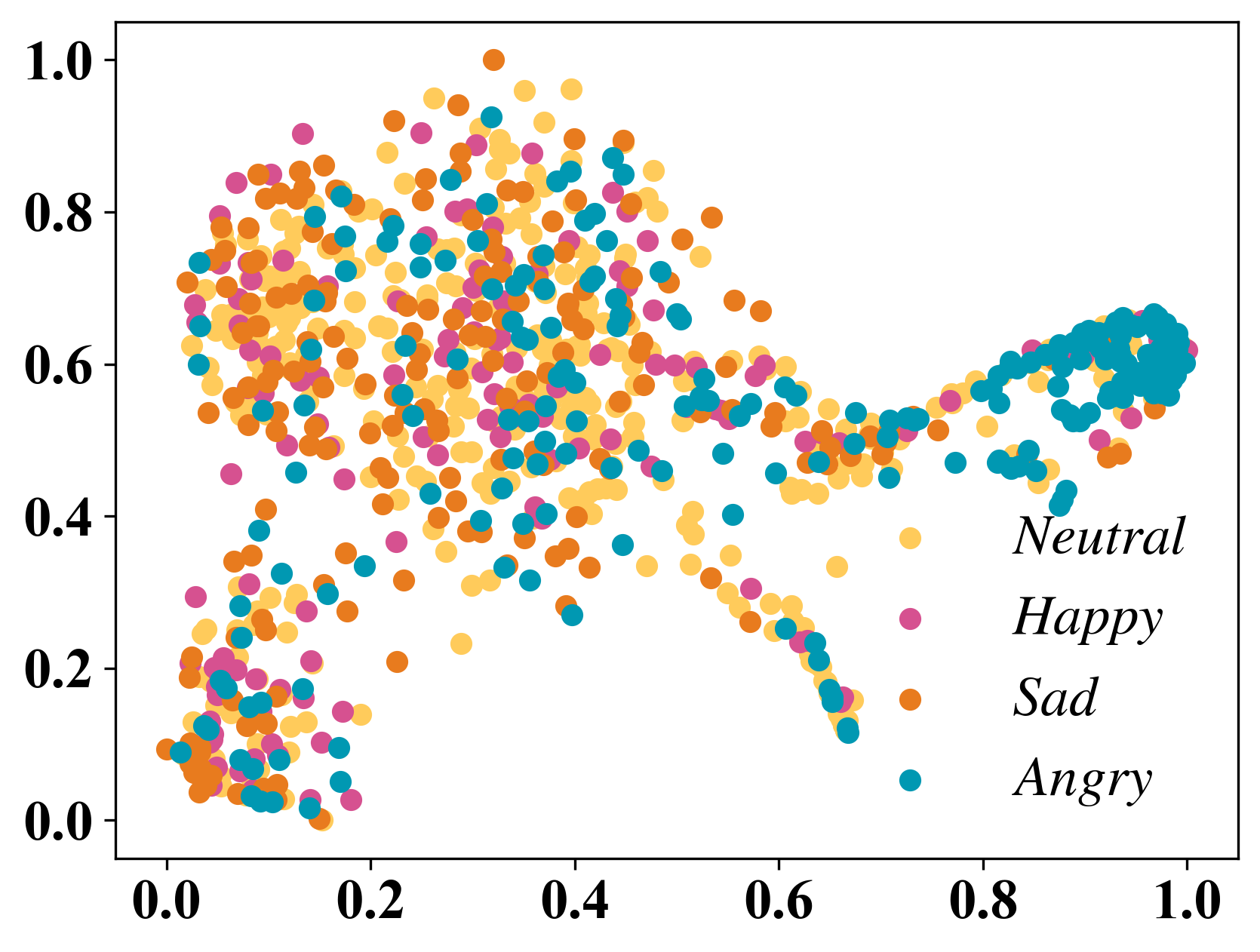}\label{fig_URRL_WO_UBR}
\caption {Visualization of MCUR and its variants on IEMOCAP dataset with MR=0.7. (a) w/o MCB-CL; (b) w/o SUGR; (c) Full MCUR.}\label{fig_URRL_VISUAL}
\end{figure*}

To better understand the impact of the MCB-CL and SUGR modules in MCUR, we visualized the representations tested on the IEMOCAP dataset with $MR=0.7$, where each sample contained only one modality. The representations are obtained from MCUR, MCUR without MCB-CL, and MCUR without the SUGR, as shown in Figure \ref{fig_URRL_VISUAL}.

Our observations reveal that removing either module leads to more scattered representations, with the absence of the SUGR module resulting in the most pronounced dispersion. This suggests that without the SUGR module, the uncertainty introduced by missing modalities is not effectively accounted for, which disrupts the clustering of representations. On the other hand, removing the MCB-CL module also leads to a less cohesive representation, likely because it eliminates the shared label information across combinations of the same modalities. When both modules are included, the resulting clusters are much more defined, highlighting the effectiveness of both modules in MCUR.

\subsection{Discussion of Limitations and Societal Impacts}
\subsubsection{Discussion of Limitation} 
Although MCUR demonstrates strong performance across various scenarios, several limitations remain. First, the dataset constraint poses a challenge. Due to the lack of datasets that truly reflect real-world multimodal conditions, our experiments mirror prior work by simulating missing modalities by randomly discarding modality representations. This approximation may not fully capture the complexity and unpredictability of real-world data loss. Second, from a model design perspective, our approach, like other KD-based methods, requires training an additional teacher model. This increases computational cost and complicates deployment in resource-limited environments. Finally, the model’s sensitivity to data variation is worth noting. Differences in background and scene characteristics across datasets, for instance, monologue settings in MOSI and MOSEI versus multi-person dialogues in IEMOCAP, as well as randomness in modality dropout, can lead to fluctuations in performance. In future work, we aim to explore models that can better handle real-world conditions with missing modalities, achieving greater efficiency, stability, and robustness without relying on simulated modality loss or additional teacher networks.

\subsubsection{Discussion of Societal Impacts}  Recognizing emotion from incomplete multimodal data improves the robustness and applicability of MER systems. However, this technology also poses potential societal risks. For instance, inferring emotions from partial, noisy or missing data may lead to inaccurate profiling, privacy breaches, or unintended surveillance without user consent.

\end{document}